\documentclass[aps,preprint, showkeys]{revtex4}
\usepackage{amsmath}
\newcommand{\be}{\begin{equation}}
\newcommand{\ee}{\end{equation}}
\newcommand{\bea}{\begin{eqnarray}}
\newcommand{\eea}{\end{eqnarray}} 
\newcommand{\ba}{\begin{array}}
\newcommand{\ea}{\end{array}}

\newcommand{\bb}{\bibitem}
\begin{document}
\begin{flushright}
\phantom{a}
\vspace{-3cm}\large
YITP-SB-16-36\\
\normalsize
\end{flushright}
\title{\bf Chiral Closed strings: Four massless states scattering amplitude}
\author{Marcelo M. Leite\footnote{Permanent Address: Laborat\'orio de F\'\i sica Te\'orica e Computacional, Departamento de F\'\i sica, Universidade Federal de Pernambuco, 
50670-901, Recife, PE, Brazil (mleite@df.ufpe.br)}} 
\author{Warren Siegel\footnote{siegel@insti.physics.sunysb.edu}}
\affiliation{{\it C N Yang Institute for Theoretical Physics, State University of New York, Stony Brook, NY 11794-3840}}

%\end{center}
\vspace{0.2cm}
\begin{abstract}
{\it We compute the scattering amplitudes of four massless states for chiral (closed) bosonic and type II 
superstrings using the Kawai-Lewellen-Tye ($KLT$) factorization method. The amplitude in the chiral bosonic case is 
identical to a field theory amplitude corresponding to the spin-$2$ 
tachyon, massless gravitational sector and massive spin-2 tardyon states of the spectrum. Chiral type II superstrings amplitude 
only possess poles associated with the massless gravitational sector. We briefly discuss the extension of the calculation 
to heterotic superstrings.}
\end{abstract}
\vspace{1cm}

\maketitle

\newpage
\section{Introduction}
\par The connection among field-theoretic scattering amplitudes and string theory amplitudes proposed in a series of papers 
by Cachazo, He and Yuan ($CHY$) \cite{CH1,CH2,CH3,CH4} shed further light in the mathematical properties of scattering 
amplitudes. Within this setting, supersymmetric gauge theories, gravity and also field theory 
with arbitrary spin in arbitrary dimensions might be better understood in principle, since the final form of the amplitude with 
just few simple terms eliminate the necessity of dealing with a lot of cancelations involved in intermediate steps.  
\par In the string-theoretic context, the analogue of the approach proposed by $CHY$ includes the scattering equations 
\cite{GM} as constraints implemented inside the amplitude through the insertion of Dirac delta functions in the vertex operators 
(see also {\it e. g.} \cite{BDTV}). A second alternative to stringy scattering amplitudes  
proposes a new gauge condition on the worldsheet which effectively yields the scattering equations after integration over 
the worldsheet variable $\bar{z}$, without the use of delta functions \cite{S1}. Recently, a combination of the two methods 
was proposed. The modification of the boundary conditions for the worldsheet fields along with the choice of conformal gauge 
originated the ``chiral'' string in its different versions: bosonic, type II and heterotic strings \cite{S2}. 
\par It is noteworthy to point out that the new chiral strings described in the present work differ from other chiral strings already discussed in the literature. It is important to emphasize their difference and similarities. Whereas the usual standard string theories have towers of massive states with higher angular momentum/spin, a curious aspect of chiral string theories is the finite number of states in the spectrum. Inspired in the work involving tensionless strings \cite{GRRA}, the first analysis which indicated only two massive states with the same absolute value but with different signs for the masses in the spectrum of the bosonic string (as well as three massless states) was carried out in \cite{HMS} in the context of BRST approach, but the model was discarded due to inconsistencies: those states have negative norm and decouple of physical scattering amplitudes. 
\par More recently, Ambitwistors chiral strings were developed \cite{MS}. In this chiral theory only massless states survive in the spectrum due to the absence of normal ordering. This happens because the $OPE$ $<XX>$ is trivial such that important operators like $e^{ik.X}$ do not acquire anomalous conformal weight, therefore not being able to compensate the anomalous conformal weight of a generic polynomial, say, in 
$\partial X^{\mu}$ for the values of $k^{2}$ on the mass-shell. In addition, all the ghosts and fields of the various Ambitwistor strings are purely holomorphic. An interesting connection among the Ambitwistor strings with tensionless and tensionful strings was assessed very recently in Ref. \cite{CT}. From now on we switch to the chiral string theory that is going to be described below with ghosts and fields possessing holomorphic and antiholomorphic components and with nontrivial OPES; see below. 
\par As mentioned above, one of the amazing features of these new closed string theories is the finite number of states in their spectrum. The chiral bosonic string possesses only three mass levels: a spin-$2$ tachyon, massless (``gravitational'') sector and a massive 
spin-$2$ ``tardyon''. The chiral type II superstring only contains the massless states, whereas the chiral heterotic superstring can 
have either tachyon or tardyon (never both) beyond the massless sector.    
\par In the present work we compute the scattering amplitude of four massless chiral string states for the bosonic and 
type II superstrings using the Kawai-Lewellen-Tye \cite{KLT} factorization of closed string amplitudes as the product of 
open string amplitudes. Remarkably, their form indicates that they can be expressed entirely as amplitudes obtained using field theory 
techniques. Furthermore, we use the factorization property to prove that they can be obtained as the product of two three-point amplitudes 
involving two external massless and one intermediate arbitrary state of the corresponding string spectrum. Using 
common aspects of bosonic, type II and heterotic strings, we construct the scattering amplitude of four massless states of the 
heterotic superstring at the massless gravitational sector. 

\section{Review of Vertex Operators and Computation of 3-Point Scattering Amplitudes}
\subsection{Closed Chiral Bosonic String}
\par The chiral bosonic string has only the matter fields $X^{\mu}(z,\bar{z})$ and conformal ghosts 
$(b(z),c(z), \tilde{b}(\bar{z}), \tilde{c}(\bar{z}))$. The operator product expansion of $X^{\mu}$ is modified (due to a Bogoliubov 
transformation in the oscillators) when compared to the standard bosonic string to 
\begin{eqnarray}\label{1}
&& <X^{\mu}(z_{i},\bar{z}_{i})X^{\nu}(z_{j},\bar{z}_{j})> = \frac{\alpha'}{2}\eta^{\mu \nu} ln\left(\frac{\bar{z}_{ij}}{z_{ij}}\right),
\end{eqnarray}
where $z_{ij}=z_{i}-z_{j}$. (The OPE for the ghosts also presents a factor of $(-1)$ in the right mover sector, namely 
$<\tilde{b}(\bar{z}_{i}), \tilde{c}(\bar{z}_{j})>=-\frac{1}{\bar{z}_{ij}}$). The modification in comparison with the usual bosonic string 
occurs only in the right-mover sector characterized by the coordinates $\bar{z}_{i}$. This can be interpreted as a change of 
sign (``flipping'') in the metric for all right-mover fields. For instance, the operator $e^{ik.X(z,\bar{z})}$ for the chiral string 
has conformal dimension $(\frac{\alpha' k^{2}}{4},-\frac{\alpha' k^{2}}{4})$. The spectrum has only 3 mass levels: a spin-$2$ 
tachyon ($M^{2}=-\frac{4}{\alpha'}$), a massless state (composed by a graviton, a massless two-form field and a massless scalar dilaton) 
and a spin-$2$ tardyon ($M^{2}=\frac{4}{\alpha'}$). 
\par The spectrum can be classified in terms of vertex operators as follows.
 
1. Tachyonic sector  ($M^{2}=-\frac{4}{\alpha'}$) whose vertex operator is given by
\begin{eqnarray}\label{2}
&& V_{-}= \bar{\epsilon}_{\mu \nu} \bar{\partial} X^{\mu} \bar{\partial} X^{\nu} e^{ik.X(z,\bar{z})}.
\end{eqnarray}

 2. Massless gravitational sector 
(includes graviton, dilaton and antisymetric tensor), with   
\begin{eqnarray}
&& V_{0}= e_{\mu \nu}\partial X^{\mu} \bar{\partial} X^{\nu} e^{ik.X(z,\bar{z})},\label{3}\\
&& e_{\mu \nu}= e_{\mu} \tilde{e}_{\nu}.\label{4}
\end{eqnarray}

3. Tardyonic sector ($M^{2}=\frac{4}{\alpha'}$) vertex operator reads
\begin{eqnarray}\label{5}
&& V_{+}= \epsilon_{\mu \nu} \partial X^{\mu} \partial X^{\nu} e^{ik.X(z,\bar{z})}.
\end{eqnarray}
\par Observe that the polarization tensor of the gravitational sector was decomposed into the direct product of two vector polarizations. 
This will help to present the scattering amplitudes in their factorized form. The location of three vertex operators in a generic 
tree-level amplitude have to be specified in the scattering of, say, $N$ states. For bosonic strings this specification consists in 
attaching the ghost factor $c(z) \tilde{c}(\bar{z})$ in the exact position of the vertex operator. For $N>3$, in addition to the above 
insertions, the remaining $(N-3)$ coordinates are integrated over $(z,\bar{z})$. 
\par Before proceeding in the evaluation of scattering amplitudes, we anticipate that the tachyonic and tardyonic states have ghost-like nature. We shall unveil this feature later, when performing the comparison using the $KLT$ computation of the four massless external states with the product of two three-point scattering amplitudes involving two external massless states (gravitational sector) and one internal tachyon or tardyon. The agreement among these results only occurs exactly if the propagator for the tachyon or tardyon includes a factor of $(-1)$, an explicit evidence of their ghost-like character. This should be contrasted with the situation when the gravitational sector (composed of graviton, a two-form field and a dilaton) is the intermediate state in a three-point scattering amplitude. In that case, the $KLT$ prescription for the four massless scattering amplitude agrees with the product of two three-point massless amplitudes without the 
need of including the factor of $(-1)$ in the propagator, corroborating the physical nature of the massless states components of the gravitational sector. 
\par We can make use of these ingredients to compute all possible 3-point scattering amplitudes. We shall restrict ourselves to a minimal 
set that are going to be useful later on. Let $A_{03} (k_{1},e_{1};k_{2},e_{2};k_{3},e_{3})$ denote the scattering amplitude for three massless states. Its evaluation yields the following factorized form:
\begin{eqnarray}
&& A_{03} (k_{1},e_{1};k_{2},e_{2};k_{3},e_{3}) = C A_{03L}A_{03R},\label{6}\\
&& A_{03L} =  e_{1 \mu_{1}} e_{2 \mu_{2}} e_{3 \mu_{3}}[\eta^{\mu_{1} \mu_{2}} k_{1}^{\mu_{3}} + \eta^{\mu_{1} \mu_{3}} k_{3}^{\mu_{2}} 
+ \eta^{\mu_{2} \mu_{3}} k_{2}^{\mu_{1}} + \frac{\alpha'}{2} k_{2}^{\mu_{1}} k_{3}^{\mu_{2}} k_{1}^{\mu_{3}} ],\label{7}\\
&& A_{03R} =  \tilde{e}_{1 \nu_{1}} \tilde{e}_{2 \nu_{2}} \tilde{e}_{3 \nu_{3}}[\eta^{\nu_{1} \nu_{2}} k_{1}^{\nu_{3}} + \eta^{\nu_{1} \nu_{3}} k_{3}^{\nu_{2}} 
+ \eta^{\nu_{2} \nu_{3}} k_{2}^{\nu_{1}} - \frac{\alpha'}{2} k_{2}^{\nu_{1}} k_{3}^{\nu_{2}} k_{1}^{\nu_{3}} ].\label{8}
\end{eqnarray} 
\par Needless to say, we employ the "transversality" condition $k_{i \mu}.e_{i}^{\mu}=0$ for the graviton vector polarization as well as 
$k_{i \mu} \epsilon_{i}^{\mu \nu}=k_{i \mu} \bar{\epsilon}_{i}^{\mu \nu}=0$ for the tachyon and tardyon tensor polarizations throughout the paper.  
\par Note that all these expressions are manifestly gauge invariant: if one of the polarization vectors $e_{i \mu_{i}}$ is replaced by the momentum $k_{i \mu_{i}}$ each amplitude above vanishes as a consequence of momentum conservation and 
the on-shell conditions. We will not be concerned with the exact value of the constants appearing in front of the 
expression of all amplitudes. Since the interesting part of any amplitude is given by its content in 
terms of momenta and polarization tensors, our carelessness concerning the overall factor will cause no trouble in the remainder of the 
discussion.
\par All states have momentum and tensor polarization (associated with each ``leg'' of an hypothetical Feynman diagram) in all states of the 
bosonic string just like in ordinary field theory. Hence we shall omit the arguments taking this fact into account explicitly in the scattering 
amplitudes. This dependence can be made explicit anytime we see fit to our discussion. We carry on this convention to the type $II$ and heterotic 
superstrings as well.
\par Analogously, the scattering amplitude for 2 massless and 1 tachyonic state denoted by $A_{021-}$ can be written in factorized form as
\begin{eqnarray}
&& A_{021-} = C A_{021- L}A_{021- R},\label{9}\\
&& A_{021- L} =  e_{1 \mu_{1}} e_{2 \mu_{2}}[\eta^{\mu_{1} \mu_{2}} -\frac{\alpha'}{2}k_{2}^{\mu_{1}}k_{1}^{\mu_{2}}],\label{10}\\
&& A_{021- R} =  \tilde{e}_{1 \nu_{1}} \tilde{e}_{2 \nu_{2}} \bar{\epsilon}_{3 \lambda \eta}[\eta^{\nu_{1} \lambda} \eta^{\eta \nu_{2}} + 
\eta^{\nu_{1} \eta} \eta^{\nu_{2} \lambda} - \frac{\alpha'}{2}(\eta^{\nu_{1} \lambda} k_{1}^{\nu_{2}} k_{2}^{\eta} \nonumber\\
&&  +  \eta^{\nu_{1} \eta} k_{1}^{\nu_{2}} k_{2}^{\lambda} + \eta^{\nu_{2} \lambda} k_{1}^{\eta} k_{2}^{\nu_{1}} + \eta^{\nu_{2} \eta} k_{1}^{\lambda} k_{2}^{\nu_{1}} 
- \eta^{\nu_{1} \nu_{2}} k_{1}^{\eta} k_{2}^{\lambda}) 
+ (\frac{\alpha'}{2})^{2} k_{2}^{\nu_{1}} k_{1}^{\nu_{2}} k_{1}^{\eta} k_{2}^{\lambda}].\label{11}
\end{eqnarray}  
\par The simplest way to check the gauge invariance of this amplitude is to look into the object $A_{021- L}$. When a 
polarization vector $e_{i \mu_{i}}$ is substituted by the momentum $k_{i \mu_{i}}$, for $i=1, 2$, using momentum conservation $k_{1}+k_{2}+k_{3} =0$ ($k_{3}$ is the intermediate tachyon momentum) with the on-shell conditions 
$k_{1}^{2}=k_{2}^{2}=0$ as well as $k_{3}^{2}= \frac{4}{\alpha'}$, $A_{021- L}$ is automatically zero. This mechanism of gauge invariance is a little different from the one with the gravitational sector as the intermediate state in the three-point amplitude. However, since the gauge invariance argument for the tachyonic sector is in complete analogy 
with the case of the scattering process of the two external massless states and one intermediate tardyon state, we let the reader figure out the gauge invariance of this three-point amplitude along the same steps described above. 
\par Next, the scattering amplitude for 2 massless and 1 tardyonic state $A_{021+}$ can be 
shown to be given by
\begin{eqnarray}
&& A_{021+} = C A_{021+ L}A_{021+ R},\label{12}\\
&& A_{021+ R} =  \tilde{e}_{1 \mu_{1}} \tilde{e}_{2 \mu_{2}}[\eta^{\mu_{1} \mu_{2}} + \frac{\alpha'}{2}k_{2}^{\mu_{1}}k_{1}^{\mu_{2}}],\label{13}
\end{eqnarray}
\begin{eqnarray}
&& A_{021+ L} =  e_{1 \nu_{1}} e_{2 \nu_{2}} \epsilon_{3 \lambda \eta} [\eta^{\nu_{1} \lambda} \eta^{\eta \nu_{2}} + 
\eta^{\nu_{1} \eta} \eta^{\nu_{2} \lambda} + \frac{\alpha'}{2}(\eta^{\nu_{1} \lambda} k_{1}^{\nu_{2}} k_{2}^{\eta}\nonumber\\
&& + \eta^{\nu_{1} \eta} k_{1}^{\nu_{2}} k_{2}^{\lambda} + 
\eta^{\nu_{2} \lambda} k_{1}^{\eta} k_{2}^{\nu_{1}} + \eta^{\nu_{2} \eta} k_{1}^{\lambda} k_{2}^{\nu_{1}} - \eta^{\nu_{1} \nu_{2}} k_{1}^{\eta} k_{2}^{\lambda}) 
+ \bigl(\frac{\alpha'}{2} \bigr)^{2} k_{2}^{\nu_{1}} k_{1}^{\nu_{2}} k_{1}^{\eta} k_{2}^{\lambda}].\label{14}
\end{eqnarray}  
\par It is worthy noting that the left-mover scattering amplitude in the case of two massless and one tachyon states does not depend on the 
tachyon polarization and involves only massless polarization vectors. The right-mover scattering of two massless excitations and one tardyon state, 
on the other hand, does not depend on the tardyon polarization. They resemble one piece of the left- and right-mover amplitude for 
three massless states. These amplitudes possess the symmetry of flipping of the sign of the metric in the right-mover sector.
\subsection{Type $II$ Chiral Superstrings}
\par We are going to focus only in the bosonic sector of the chiral type $II$ (valid for both $IIA$ and $IIB$) superstrings in the 
Ramond-Neveu-Schwarz ($RNS$) covariant formulation. The chiral boundary conditions produce a spectrum that contains only massless fields. 
The scalar tachyon is projected out using the $GSO$ projection. The total field content in the Neveu-Schwarz ($NS$) bosonic sector is 
composed by left-mover and right-mover worldsheet fermions $\psi^{\mu}(z)$ and $\tilde{\psi}^{\nu}(\bar{z})$, as well as the bosonic 
fields $X^{\mu}(z,\bar{z})$. The $OPE$ of the worldsheet fermions are $<\psi^{\mu}(z_{i}) \psi^{\nu}(z_{j})>= \frac{\eta^{\mu \nu}}{z_{ij}}$ and 
$<\tilde{\psi}^{\mu}(\bar{z}_{i}) \tilde{\psi}^{\nu}(\bar{z}_{j})> = -\frac{\eta^{\mu \nu}}{\bar{z}_{ij}}$. Furthermore, type $II$ superstrings require 
vertex operators which are multiplied by powers of $e^{\phi(z)}$ and $e^{\tilde{\phi}(\bar{z})}$ (called pictures, a residual effect of the 
fermionization of the bosonic ghosts). These vertex operators 
naturally produce conformally invariant scattering amplitudes. The worldsheet fermionic part of the massless NS-NS vertex operators comes from 
$\psi^{\mu}(z)$, $\tilde{\psi}^{\mu}(\bar{z})$, and $e^{-\phi},e^{-\tilde{\phi}}$. The $OPE$s $<\phi(z_{i}) \phi(z_{j})>= ln z_{ij}$, 
$<\tilde{\phi}(\bar{z}_{i}) \tilde{\phi}(\bar{z}_{j})>= ln \bar{z}_{ij}$ are given for the sake of completeness.
\par We employ the standard choice for the pictures of vertex operators. There are two types of vertex operators, namely, those in the 
$(0,0)$ picture and in the $(-1,-1)$ picture. The ones in the $(-1,-1)$ picture correspond to each of the end states (the ``in'' and ``out'' 
states); the intermediate vertex operators are said to be in the $(0,0)$ picture. At fixed positions, the latter are attached to $c(z)\tilde{c}(\bar{z})$ factors, whereas if the position is not specified the factor $\int d^{2}z$ replace the product of $c$ ghost factors. 
\par Taking into account these observations, the vertex operators for the massless states 
are given by the following expressions:
\begin{eqnarray}
&& \mathcal{V}^{(-1,-1)} = e_{\mu \nu}e^{-\phi} e^{-\tilde{\phi}}\psi^{\mu}\tilde{\psi}^{\nu}e^{ik.X(z,\bar{z})},\label{15}\\
&& \mathcal{V}^{(0,0)} = e_{\mu \nu} (-\frac{2}{\alpha'})(i \partial X^{\mu} + \frac{\alpha'}{2} k.\psi \psi^{\mu})
(-i \bar{\partial}X^{\nu} + \frac{\alpha'}{2} k.\tilde{\psi} \tilde{\psi}^{\mu})e^{ik.X(z,\bar{z})}.\label{16}
\end{eqnarray}  
\par Using these vertex operators, we get the following scattering amplitude for three massless states:
\begin{eqnarray}
&& \mathcal{A}_{03} = C \mathcal{A}_{03L} \mathcal{A}_{03R}, \label{17}\\
&& \mathcal{A}_{03L} = e_{1 \mu_{1}} e_{2 \mu_{2}} e_{3 \mu_{3}}(k_{1}^{\mu_{3}} \eta^{\mu_{1} \mu_{2}} + k_{3}^{\mu_{2}} \eta^{\mu_{1} \mu_{3}} 
+ k_{2}^{\mu_{1}} \eta^{\mu_{2} \mu_{3}}),\label{18}\\
&& \mathcal{A}_{03R} = \tilde{e}_{1 \nu_{1}}\tilde{e}_{2 \nu_{2}}\tilde{e}_{3 \nu_{3}} (k_{1}^{\nu_{3}} \eta^{\nu_{1} \nu_{2}} + k_{3}^{\nu_{2}} \eta^{\nu_{1} \nu_{3}} + k_{2}^{\nu_{1}} \eta^{\nu_{2} \nu_{3}}).\label{19}
\end{eqnarray} 
 These  expressions are simpler than those from the bosonic string. The amplitude is just equal to the bosonic one without the 
$O(\alpha')$ correction (cubic in the momenta). In addition, gauge invariance is manifest as well, by following the same reasoning of the gauge invariance presented to the scattering of three massless states in the bosonic string.
\subsection{Heterotic Chiral Superstring}
\par For the heterotic superstrings, two choices can be made: either the left-mover sector is bosonic and the right-mover sector is 
supersymmetric or the left-mover sector is supersymmetric and the right-mover sector is bosonic. In the first situation, only heterotic 
spin-$2$ tardyon and massless modes are present. In the second scenario, the spectrum contains only massless and spin-$2$ tachyon states.
\par Consider the heterotic superstring with bosonic left-movers $X^{\mu}$ ($\mu=0,..., 9$) and Majorana-Weyl worldsheet fermions 
$\lambda^{A}$ ($A=1,...,32$ represents quantum internal symmetry groups of the fermions, for instance $SO(32), E_{8} \times E_{8}$, 
etc). We could also have considered the other possibility where the tachyon is present. However, we shall focus only on the heterotic 
string with the tardyon in the remainder. 
\par  The heterotic superstring has a modified type of $GSO$ projection: identical to the open superstring in the right-mover sector and 
restricted to an even number of creation operators of the internal symmetry in the left-mover sector. The bosonic massless heterotic states 
come from the NS sector: no fermionic excitations from the internal symmetry are present. The 
state-vertex operator correspondence for chiral heterotic superstring translates ``$in$'' and ``$out$'' states into vertex operators in 
the $(0,-1)$ picture
\begin{equation}
\mathcal{V}_{h}^{(0,-1)}= e_{\mu \nu} e^{-\tilde{\phi}} :\partial X^{\mu} \tilde{\psi}^{\nu} e^{ik.X(z,\bar{z})}:,\label{20}
\end{equation}
whereas intermediate states correspond to vertex operators in the $(0,0)$ picture: 
\begin{equation}
\mathcal{V}_{h}^{(0,0)}= e_{\mu \nu} (-\frac{2}{\alpha'}):i \partial X^{\mu}(-i \partial X^{\nu} 
+ \frac{\alpha'}{2} k. \tilde{\psi} \tilde{\psi}^{\nu}) e^{ik.X(z,\bar{z})}: .\label{21}
 \end{equation}
\par The recipe to compute scattering amplitudes is similar to the description given for the type II superstring. For instance, the 3-point 
scattering amplitude corresponding to three massless states results in the expression 
\begin{eqnarray} 
&& \mathcal{A}_{03h}= C \mathcal{A}_{03Lh} \mathcal{A}_{03Rh},\label{22}\\
&& \mathcal{A}_{03Lh}=  e_{1 \mu_{1}} e_{2 \mu_{2}} e_{3 \mu_{3}}[\eta^{\mu_{1} \mu_{2}} k_{1}^{\mu_{3}} + \eta^{\mu_{1} \mu_{3}} k_{3}^{\mu_{2}} 
+ \eta^{\mu_{2} \mu_{3}} k_{2}^{\mu_{1}} + \frac{\alpha'}{2} k_{2}^{\mu_{1}} k_{3}^{\mu_{2}} k_{1}^{\mu_{3}} ],\label{23}\\
&& \mathcal{A}_{03Rh}=  \tilde{e}_{1 \nu_{1}}\tilde{e}_{2 \nu_{2}}\tilde{e}_{3 \nu_{3}} (k_{1}^{\nu_{3}} \eta^{\nu_{1} \nu_{2}} + k_{3}^{\nu_{2}} \eta^{\nu_{1} \nu_{3}} + k_{2}^{\nu_{1}} \eta^{\nu_{2} \nu_{3}}).\label{24}
\end{eqnarray} 
\par It is not dificult to see that the left-mover contribution is identical to the bosonic amplitude and the right-mover one is equal 
to that from the type $II$ superstring for the scattering of three massless states. For later convenience, from now on we drop the index 
notation in the above equations and adopt scalar products instead. From our previous discussion, it is obvious that gauge invariance holds here as well. 
\section{\bf Four Massless States Scattering Amplitude}
\subsection{\bf Chiral Bosonic Closed String} 
\par We now use the $KLT$ \cite{KLT} factorization method for the chiral bosonic string in the evaluation of the four-point 
scattering amplitude involving massless states. We shall fix the position of three vertex operators at points $z_{1}=0$, 
$z_{3} =1$ and $z_{4}=\infty$ and let the vertex operator at the position $z_{2}$ to be integrated. We postpone the 
usage of the explicit chosen values of $z_{1}, z_{3}, z_{4}$ until the actual computation of the integral over the holomorphic and antiholomorphic variables; see below. The explicit 
expression of the amplitude in the sphere is given by ($G_{4}^{S^{2}} \equiv G_{4}$)
\begin{eqnarray}
&& G_{4} = A \int d^{2}z_{2} <c(z_{1})\tilde{c}(\bar{z}_{1}) V_{0}(z_{1},\bar{z}_{1}) V_{0}(z_{2},\bar{z}_{2}) c(z_{3}) 
\tilde{c}(\bar{z}_{3}) V_{0}(z_{3},\bar{z}_{3}) \;\; \times \nonumber\\
&& \;\; c(z_{4}) \tilde{c}(\bar{z}_{4}) V_{0}(z_{4},\bar{z}_{4})>,\label{25}
\end{eqnarray}
where $A$ is a constant. Now we employ the $OPE$ of the fields $X^{\mu}(z_{i},\bar{z}_{i})$ as well as the result 
$<c(z_{1})\tilde{c}(\bar{z}_{1})c(z_{3}) 
\tilde{c}(\bar{z}_{3}) c(z_{4}) \tilde{c}(\bar{z}_{4})> = C_{S^{2}}^{g} z_{13} \bar{z}_{13} z_{14} \bar{z}_{14} z_{24} \bar{z}_{24}$ ($C_{S^{2}}^{g}$ is a constant which will be absorbed in a redefinition of the overall constant $A$). Moreover, it is easy to prove that the 
Koba-Nielsen factor for the chiral bosonic string yields the expectation value in the sphere
\begin{eqnarray}\label{26}
&& <e^{ik.X(z_{1},\bar{z}_{1})}e^{ik.X(z_{2},\bar{z}_{2})}e^{ik.X(z_{3},\bar{z}_{3})}e^{ik.X(z_{4},\bar{z}_{4})}> = 
[iC_{S^{2}}^{X}(2 \pi)^{26} \delta^{26}(\sum_{i=1}^{4}k_{i})] \Bigl(\frac{z_{12}}{\bar{z}_{12}}\Bigr)^{\frac{\alpha'}{2} k_{1}.k_{2}} 
\Bigl(\frac{z_{13}}{\bar{z}_{13}}\Bigr)^{\frac{\alpha'}{2} k_{1}.k_{3}} \nonumber\\
&& \Bigl(\frac{z_{14}}{\bar{z}_{14}}\Bigr)^{\frac{\alpha'}{2} k_{1}.k_{4}} 
\Bigl(\frac{z_{23}}{\bar{z}_{23}}\Bigr)^{\frac{\alpha'}{2} k_{2}.k_{3}} \Bigl(\frac{z_{24}}{\bar{z}_{24}}\Bigr)^{\frac{\alpha'}{2} k_{2}.k_{4}}
\Bigl(\frac{z_{34}}{\bar{z}_{34}}\Bigr)^{\frac{\alpha'}{2} k_{3}.k_{4}}.
\end{eqnarray}
As before, the constant $iC_{S^{2}}^{X}(2 \pi)^{26}$ is also going to be absorbed in the redefinition of $A$ above. Since momentum 
conservation will be used over an over again in all computations, we do not bother to write the delta function in the expression for 
the amplitude. 
\par The Mandelstam variables using $KLT$ conventions are 
defined by $s=-(k_{1}+k_{2})^{2}$, $t=-(k_{2}+k_{3})^{2}$ and $u=-(k_{2}+k_{4})^{2}$. When we use the value of these variables on-shell 
where $k_{i}^{2}=0$ for $i=1,...,4$, we obtain the constraint associated with the scattering of four massless string states $s+t+u=0$. Now, 
we turn our attention to the actual computation of the integral. The basic steps are: first insert the value of the several $OPE$'s 
appearing above along with the aforementioned results. Second, employ the values $z_{1}=0$, 
$z_{3} =1$ and $z_{4}=\infty$. One is left with the task of computing the integral over the variables 
$z_{2},\bar{z}_{2}$ which can be performed independently in the $KLT$ approach. The $KLT$ method considers the 
integration region in the interval $(0,1)$ in the left-mover variable $z_{2}$ and the interval $(1,\infty)$ in the right-mover variable $\bar{z}_{2}$, thanks to the monodromy properties of the string correlators, therefore relating open string amplitudes with closed string amplitudes. The overall result produces closed string amplitudes as the factorization (product) of open string amplitudes.
\par The first integral is a function of the Mandelstam variables $(s,t)$. After performing the integral one gets an overall factor 
of (-1) along with products of Gamma functions involving all Mandelstam variables. In the second integral, the integration region can 
be transformed into the interval $(0,1)$ after a M$\ddot{o}$bius transformation, producing a function of ($t,u$). After integration, 
all contributions to the amplitude receive a common monodromy factor $sin\Bigl(\frac{\pi \alpha' t}{4}\Bigr)$ multiplied by Gamma 
functions as well. Using the identity $sin(\pi b)= \frac{\pi}{\Gamma(b)\Gamma(1-b)}$, the overall factor multiplying the right-mover 
and left-mover contributions combine together in the expression of the complete amplitude producing a result in terms of $\Gamma$ 
functions that is equivalent to the product of two open string amplitudes. 
\par When we perform the multiplication of these two contributions we obtain a product of two factors, each one with a different origin.
The first one results from  using the several identities involving Gamma functions as explained 
above and produces the factor $\frac{\pi}{stu}$. This result is remarkable. We refer the interested reader to Ref. \cite{S2} in order to see explicitly this non-trivial feature at work. The second one consists of a sum of combinations of polynomials in $s,t,u$ along with 
poles in the tachyon and in the tardyon positions in the $s-, t-$ and $u-$ channels. 
\par The product of these two factors results in the sum of singular terms which contains single poles and other contributions that 
can be transformed into simple poles. The simplification in the latter can be achieved using the identities 
$\frac{1}{(1+x)(1-x)}=\frac{1}{2(1-x)}+\frac{1}{2(1+x)}$; $\frac{1}{x(1+x)(1-x)}=\frac{1}{x} + \frac{1}{2(1-x)}-\frac{1}{2(1+x)}$, 
$\frac{z}{xy}=-1/x -1/y$ and $\frac{z}{(1+x)(1-y)}=\frac{1}{(1+x)}-\frac{1}{(1-y)}$ with $z=-x-y$. Using these facts, after a tedious 
but straightforward computation we get to the expression for the complete amplitude written as            
\begin{eqnarray}
&& G_{4} = B (-1)\Bigr[\frac{H_{0s}}{\frac{\alpha' s}{4}}  + \frac{H_{0t}}{\frac{\alpha' t}{4}} 
+ \frac{H_{0u}}{\frac{\alpha' u}{4}}  + \frac{H_{-s}}{1+\frac{\alpha' s}{4}} + \frac{H_{-t}}{1+\frac{\alpha' t}{4}} + \frac{H_{-u}}{1+\frac{\alpha' u}{4}} \nonumber\\
&& + \frac{H_{+s}}{1 - \frac{\alpha' s}{4}} 
+ \frac{H_{+t}}{1 - \frac{\alpha' t}{4}} + \frac{H_{+u}}{1 - \frac{\alpha' s}{4}} \Bigl],\label{27}
\end{eqnarray}
where $B$ is a constant. The form of this expression is very similar to a field theory amplitude: there are massless poles, ``tachyonic'' poles 
and ``tardyonic'' poles in the $s-$, $t-$ and $u-$channel, respectively. The coefficient of the massless poles in the $s-$, $t-$ and $u-$channels 
are given by 
\begin{eqnarray}
&& H_{0s} = (\frac{\alpha'}{4})^{2} tu P_{s} \bar{P}_{s} - \frac{\alpha'}{2}\Biggl(\frac{\alpha' t}{4}\Bigl(P_{s} \bar{C}_{t} 
+ C_{t} \bar{P}_{s}\Bigr) + \frac{\alpha' u}{4}\Bigl(P_{s} \bar{C}_{u} + C_{u} \bar{P}_{s}\Bigr)\Biggr)\nonumber\\
&&  - \frac{\alpha^{' 2}}{4}(C_{u} - C_{t})(\bar{C}_{u} - \bar{C}_{t}),\label{28}\\
&& H_{0t} = (\frac{\alpha'}{4})^{2} su P_{t} \bar{P}_{t} - \frac{\alpha'}{2}\left(\frac{\alpha' s}{4}\Bigl(P_{t} \bar{C}_{s} 
+ C_{s} \bar{P}_{t}\Bigr) + \frac{\alpha' u}{4}\Bigl(P_{t} \bar{C}_{u} + C_{u} \bar{P}_{t}\Bigr)\right)\nonumber\\
&&  - \frac{\alpha^{' 2}}{4}(C_{u} - C_{s})(\bar{C}_{u} - \bar{C}_{s}),\label{29}\\
&& H_{0u} = (\frac{\alpha'}{4})^{2} st P_{u} \bar{P}_{u} - \frac{\alpha'}{2}\left(\frac{\alpha' s}{4}\Bigl(P_{u} \bar{C}_{s} 
+ C_{s} \bar{P}_{u}\Bigr) + \frac{\alpha' t}{4}\Bigl(P_{u} \bar{C}_{t} + C_{t} \bar{P}_{u}\Bigr)\right)\nonumber\\
&&  - \frac{\alpha^{' 2}}{4}(C_{s} - C_{t})(\bar{C}_{s} - \bar{C}_{t}),\label{30}
\end{eqnarray}
where the quantitities $C_{s}$, $\bar{C}_{s}$, $C_{t}$ and $\bar{C}_{t}$ are defined by
\begin{eqnarray}
&& C_{s} (e) = k_{4}.e_{1}  k_{2}.e_{3} e_{2}.e_{4} +  k_{3}.e_{2} k_{1}.e_{4} e_{1}.e_{3} + k_{3}.e_{1} k_{2}.e_{4} e_{2}.e_{3} + k_{4}.e_{2} k_{1}.e_{3} e_{1}.e_{4} ;\label{31}\\
&& \bar{C}_{s} (\tilde{e}) = C_{s}(\tilde{e});\label{32}\\
&& C_{t} (e) = k_{1}.e_{2} k_{3}.e_{4} e_{1}.e_{3} + k_{2}.e_{1} k_{4}.e_{3} e_{2}.e_{4} + k_{3}.e_{1} k_{4}.e_{2} e_{3}.e_{4} + k_{1}.e_{3} k_{2}.e_{4} e_{1}.e_{2}\nonumber\\
&& - \frac{\alpha'}{2} \Bigl(k_{3}.e_{1} k_{4}.e_{2}(k_{1}.e_{3} k_{1}.e_{4} + k_{2}.e_{3} k_{2}.e_{4}) -  (k_{3}.e_{1} k_{1}.e_{2} + k_{2}.e_{1} k_{4}.e_{2}) k_{1}.e_{3}  
k_{2}.e_{4}\Bigr);\label{33}
\end{eqnarray}
\begin{eqnarray}
&& \bar{C}_{t} (\tilde{e}) = k_{1}.\tilde{e}_{2} k_{3}.\tilde{e}_{4} \tilde{e}_{1}.\tilde{e}_{3} + k_{2}.\tilde{e}_{1} k_{4}.\tilde{e}_{3} \tilde{e}_{2}.\tilde{e}_{4} + k_{3}.\tilde{e}_{1} k_{4}.\tilde{e}_{2} \tilde{e}_{3}.\tilde{e}_{4} + k_{1}.\tilde{e}_{3} k_{2}.\tilde{e}_{4} \tilde{e}_{1}.\tilde{e}_{2}\nonumber\\
&& + \frac{\alpha'}{2} \Bigl(k_{3}.\tilde{e}_{1} k_{4}.\tilde{e}_{2}(k_{1}.\tilde{e}_{3} k_{1}.\tilde{e}_{4} + k_{2}.\tilde{e}_{3} k_{2}.\tilde{e}_{4}) -  (k_{3}.\tilde{e}_{1} k_{1}.\tilde{e}_{2} + k_{2}.\tilde{e}_{1} k_{4}.\tilde{e}_{2}) k_{1}.\tilde{e}_{3}  k_{2}.\tilde{e}_{4}\Bigr). \label{34}
\end{eqnarray}
These equations suggest the simpler notation $C_{t}(e)= C_{0t}(e) -\frac{\alpha'}{2}C_{1t}(e), 
\bar{C}_{t}(\tilde{e})= C_{0t}(\tilde{e}) + \frac{\alpha'}{2}C_{1t}(\tilde{e})$. Thus, we can write as well 
$C_{u}(e) = C_{0u}(e) - \frac{\alpha'}{2} C_{1u}(e), \bar{C}_{u} (\tilde{e}) = C_{0u} (\tilde{e}) + \frac{\alpha'}{2} C_{1u}(\tilde{e})$, 
where 
\begin{eqnarray}
&& C_{0u}(e) = k_{2}.e_{3} k_{1}.e_{4} e_{1}.e_{2} + k_{4}.e_{1} k_{3}.e_{2} e_{3}.e_{4} + k_{2}.e_{1} k_{3}.e_{4} e_{2}.e_{3} +  k_{1}.e_{2}  k_{4}.e_{3} e_{1}.e_{4};\label{35}\\
&& C_{1u}(e)= (k_{3}.e_{1} k_{3}.e_{2} + k_{4}.e_{1} k_{4}.e_{2}) k_{2}.e_{3} k_{1}.e_{4} - k_{4}.e_{1} k_{3}.e_{2}(k_{4}.e_{3} k_{1}.e_{4} + k_{2}.e_{3} k_{3}.e_{4}). \label{36} 
\end{eqnarray}
\par The analogue of these objects appeared previously in the $KLT$ treatment of the usual bosonic string in the kinetic term $K^{(ss)}$ there. In 
particular, the combinations $sC_{0s}+tC_{0t}+uC_{0u}$ are contained in $K^{(ss)}$ from \cite{KLT}. 
\par Similarly, the quantities corresponding to a piece of three massless states scattering amplitude, also present in the amplitudes of two 
massless and one massive state, seem to be the building blocks of all scattering amplitudes. They are defined by
\begin{eqnarray}
&& P_{s}= \Bigl(e_{1}.e_{2} -\frac{\alpha' }{2} k_{2}.e_{1} k_{1}.e_{2}\Bigr)\Bigl(e_{3}.e_{4} -\frac{\alpha' }{2} k_{4}.e_{3} k_{3}.e_{4}\Bigr);\label{37}\\
&& \bar{P}_{s}= \Bigl(\tilde{e}_{1}.\tilde{e}_{2} +\frac{\alpha'}{2} k_{2}.\tilde{e}_{1} k_{1}.\tilde{e}_{2}\Bigr)\Bigl(\tilde{e}_{3}.\tilde{e}_{4} + \frac{\alpha'}{2} k_{4}.\tilde{e}_{3} k_{3}.\tilde{e}_{4}\Bigr);\label{38}\\
&& P_{t}= \Bigl(e_{2}.e_{3} -\frac{\alpha'}{2} k_{3}.e_{2} k_{2}.e_{3}\Bigr)\Bigl(e_{1}.e_{4} -\frac{\alpha' }{2} k_{4}.e_{1} k_{1}.e_{4}\Bigr);\label{39}\\
&& \bar{P}_{t}= \Bigl(\tilde{e}_{2}.\tilde{e}_{3} +\frac{\alpha'}{2} k_{3}.\tilde{e}_{2} k_{2}.\tilde{e}_{3}\Bigr)\Bigl(\tilde{e}_{1}.\tilde{e}_{4} + \frac{\alpha'}{2} k_{4}.\tilde{e}_{1} k_{1}.\tilde{e}_{4}\Bigr);\label{40}\\
&& P_{u}= \Bigl(e_{1}.e_{3} -\frac{\alpha'}{2} k_{3}.e_{1} k_{1}.e_{3}\Bigr)\Bigl(e_{2}.e_{4} -\frac{\alpha' }{2} k_{4}.e_{2} k_{2}.e_{4}\Bigr);\label{41}\\
&& \bar{P}_{u}= \Bigl(\tilde{e}_{1}.\tilde{e}_{3} +\frac{\alpha'}{2} k_{3}.\tilde{e}_{1} k_{1}.\tilde{e}_{3}\Bigr)\Bigl(\tilde{e}_{2}.\tilde{e}_{4} + \frac{\alpha'}{2} k_{4}.\tilde{e}_{2} k_{2}.\tilde{e}_{4}\Bigr).\label{42}
\end{eqnarray} 
\par Consider the massive levels as intermediate states, namely, the (-) ``tachyonic'' and (+) ``tardyonic'' sectors. 
The coefficient of the poles in the amplitude at different channels are given in terms of our definitions above as 
\begin{eqnarray}
&& H_{-s} = P_{s}\left[-\frac{1}{2}(\frac{\alpha'}{4})^{2} tu \bar{P}_{s} + \frac{\alpha'}{2}\Bigl(\frac{\alpha' t}{4} \bar{C}_{t} 
+ \frac{\alpha' u}{4} \bar{C}_{u} -\bar{C}_{s} \Bigr) + \bar{P}_{u} + \bar{P}_{t} \right],\label{43}\\
&& H_{-t} = P_{t}\left[-\frac{1}{2}(\frac{\alpha'}{4})^{2} su \bar{P}_{s} + \frac{\alpha'}{2}\Bigl(\frac{\alpha' s}{4} \bar{C}_{s} 
+ \frac{\alpha' u}{4} \bar{C}_{u} -\bar{C}_{t} \Bigr) + \bar{P}_{s} + \bar{P}_{u} \right],\label{44}\\
&& H_{-u} = P_{u}\left[-\frac{1}{2}(\frac{\alpha'}{4})^{2} ts \bar{P}_{u} + \frac{\alpha'}{2}\Bigl(\frac{\alpha' s}{4} \bar{C}_{s} 
+ \frac{\alpha' t}{4} \bar{C}_{t} -\bar{C}_{u} \Bigr) + \bar{P}_{s} + \bar{P}_{t} \right],\label{45}\\
&& H_{+s} = \bar{P}_{s}\left[\frac{1}{2}(\frac{\alpha'}{4})^{2} tu P_{s} - \frac{\alpha'}{2}\Bigl(\frac{\alpha' t}{4} C_{t} 
+ \frac{\alpha' u}{4} C_{u} + C_{s} \Bigr) - P_{u} - P_{t} \right],\label{46}\\
&& H_{+t} = \bar{P}_{t}\left[\frac{1}{2}(\frac{\alpha'}{4})^{2} su P_{t} - \frac{\alpha'}{2}\Bigl(\frac{\alpha' s}{4} C_{s} 
+ \frac{\alpha' u}{4} C_{u} + C_{t} \Bigr) - P_{s} - P_{u} \right],\label{47}
\end{eqnarray}
\begin{eqnarray}
&& H_{+u} = \bar{P}_{u}\left[\frac{1}{2}(\frac{\alpha'}{4})^{2} ts P_{u} - \frac{\alpha'}{2}\Bigl(\frac{\alpha' s}{4} C_{s} 
+ \frac{\alpha' t}{4} C_{t} + C_{u} \Bigr) - P_{t} - P_{s} \right].\label{48}
\end{eqnarray}
\par The watchful reader might have noticed that although the amplitudes are $stu$ symmetric in their form, they are not 
explicitly $stu$ symmetric. This does not happen only in the gravitational sector, but also in the tachyonic and tardyonic sectors. The main reason is that whereas $C_{s}, \bar{C}_{s}$ do not possess $O(\alpha')$ corrections, 
$C_{t}, \bar{C}_{t}$, $C_{u}, \bar{C}_{u}$ {\it do possess them}. At $O(\alpha'^{0})$ the $stu$ symmetry is obvious for these objects, but at $O(\alpha')$ this is not so. Since these objects make part of the amplitudes, the 
potential violation in this symmetry comes from the contribution involving $C_{s}, C_{t},C_{u}$ and their counterparts in the right-mover sector. We provide an explicit non-trivial proof of the $stu$ symmetry of the scattering amplitudes at the residues of the poles in each channel to all sectors in the appendix.
\subsection{Type II Superstrings}
\par Except for the addition of the $OPE$s involving worldsheet fermions and chiral bosons, the scattering amplitude for type II 
superstrings follows the same general pattern as described for the bosonic string. Using the $KLT$ technique just as before, we 
get to  
\begin{eqnarray}
&& \mathcal{G}_{4} = B (-1)\Bigl[\frac{\mathcal{H}_{s}}{\frac{\alpha' s}{4}}  + \frac{\mathcal{H}_{t}}{\frac{\alpha' t}{4}} 
+ \frac{\mathcal{H}_{u}}{\frac{\alpha' u}{4}}\Bigr]. \label{49} 
\end{eqnarray}
where 
\begin{eqnarray}
&& \mathcal{H}_{s} = (\frac{\alpha'}{4})^{2} tu \mathcal{P}_{s} \bar{\mathcal{P}}_{s} 
- \frac{\alpha'}{2}\Bigl(\frac{\alpha' t}{4}\mathcal{P}_{s} \bar{C}_{0t} 
+ \frac{\alpha' u}{4} \mathcal{P}_{s} \bar{C}_{0u}\Bigr) \nonumber\\
&& - \frac{\alpha^{' 2}}{4}(C_{0u} - C_{0t})(\bar{C}_{0u} - \bar{C}_{0t}),\label{50}\\
&& \mathcal{H}_{t} = (\frac{\alpha'}{4})^{2} su \mathcal{P}_{t} \bar{\mathcal{P}}_{t} 
- \frac{\alpha'}{2}\Bigl(\frac{\alpha' s}{4} \mathcal{P}_{t} \bar{C}_{s} 
+ \frac{\alpha' u}{4} \mathcal{P}_{t} \bar{C}_{0u}\Bigr) \nonumber\\
&& - \frac{\alpha^{' 2}}{4}(C_{s} - C_{0u})(\bar{C}_{s} - \bar{C}_{0t}),\label{51}\\
&& \mathcal{H}_{u} = (\frac{\alpha'}{4})^{2} st \mathcal{P}_{u} \bar{\mathcal{P}}_{u} 
- \frac{\alpha'}{2}\Bigl(\frac{\alpha' s}{4} \mathcal{P}_{u} \bar{C}_{s} 
+ \frac{\alpha' t}{4}\mathcal{P}_{u} \bar{C}_{0t}\Bigr) \nonumber\\
&& - \frac{\alpha^{' 2}}{4}(C_{s} - C_{0t})(\bar{C}_{s} - \bar{C}_{0t}).\label{52}
\end{eqnarray}
Here there is no $\alpha'$ corrections in all quantities previously defined for the bosonic string. Their supersymmetric counterparts are 
defined with the special letters. For instance,  $\mathcal{P}_{s}=e_{1}.e_{2} e_{3}.e_{4}$, 
$\bar{\mathcal{P}}_{s}=\tilde{e}_{1}.\tilde{e}_{2} \tilde{e}_{3}.\tilde{e}_{4}$, etc. Consequently, the amplitude is manifestly 
$stu$ symmetric and is much simpler than in the bosonic string. 
\par It is easy to compare the results obtained so far with the heterotic string with a tardyon in its spectrum: in the left-mover sector 
it has the same structure of the bosonic string whereas its right-mover has the same structure of the type II string above. For example, if 
we attach a subscript $h$ for the heterotic string, we find that $P_{hs}=P_{s}$, $\bar{P}_{hs}=\bar{\mathcal{P}}_{s}$, etc. We can also write, 
$C_{ht}=C_{t}$, $\bar{C}_{ht}=\bar{C}_{0t}$ and so on. The factorization of the amplitudes in the $KLT$ representation permits the 
computation of the four-point scattering amplitude as well. 
\par However, we can obtain further insight in the amplitude structure using 
the factorization property in comparison with the results obtained from the product of two three arbitrary states scattering amplitudes (with two external massless states). We study 
this aspect next in order to construct the amplitude for the heterotic string without performing any integral.    
\section {Relationship with the product of two 3-point amplitudes} 
\subsection{Chiral Type II superstrings}
\par Since chiral type II superstrings only possess massless states, they are much easier to analyze. There is only a massless as intermediate 
state in the three-point function. So one multiplies the two three-point amplitudes given previously. Consider only the $s$-channel 
for the sake of simplicity. Define the four-point scattering amplitude constructed out of the product of two three-point amplitudes as 
$\tilde{\mathcal{G}}_{4s}$. In the construction of this amplitude, we include explicitly the Dirac delta functions. (We shall neglect 
the overall delta function in the final result just as we did before; see below.) It can be written schematically in the standard form
\begin{eqnarray}
&& \tilde{\mathcal{G}}_{4s} = A \int \frac{d^{10}k}{(2 \pi)^{10}} \frac{[(2 \pi)^{10}]^{2}\delta^{10}(-k+k_{1}+k_{2})\delta^{10}(k+k_{3}+k_{4})}{-k^{2}+i \epsilon} \nonumber\\
&& \sum_{polarizations} \mathcal{A}_{03}(k_{1},e_{1};k_{2},e_{2}; -k, e) \mathcal{A}_{03}(k,e;k_{3},e_{3};k_{4},e_{4}). \label{53}
\end{eqnarray}
\par The integration over $k$ yields a trivial 
product of  an overall Dirac delta function (involving the sum of all externa momenta) and the Mandelstam variable $s$ in the denominator. 
Just as before, we neglect the delta function in the resulting amplitude but take implicitly momentum conservation in all steps of the 
computation. The same systematics apply to the other channels as well with the appropriate modifications.
\par Recalling that the intermediate massless state has polarization decomposed in terms of vector polarizations as 
$e_{\mu \nu}= e_{\mu}\tilde{e}_{\nu}$, the three-point amplitudes in the two sides can be attached to each other using the 
projectors (or completeness relation for the vector polarizations) 
$\sum_{e} e_{\lambda_{1}} e_{\lambda_{2}}= \sum_{\tilde{e}} \tilde{e}_{\lambda_{1}} \tilde{e}_{\lambda_{2}} = \eta_{\lambda_{1} \lambda_{2}}$. Thus, 
it is a simple task to obtain the following value for the amplitude 
\begin{eqnarray}
&& \mathcal{\tilde{G}}_{4s}= \frac{B}{\frac{\alpha' s}{4}} \Bigl[\bigl(\frac{\alpha' u}{4}\bigr)^{2} \mathcal{P}_{s} 
\bar{\mathcal{P}}_{s} 
+ \frac{\alpha'}{2} \bigl(\frac{\alpha' u}{4}\bigr) [\mathcal{P}_{s}(\bar{C}_{0u}-\bar{C}_{0t}) + \bar{\mathcal{P}}_{s}(C_{0u}-C_{0t})]\nonumber\\
&& + \bigl(\frac{\alpha'}{2}\bigr)^{2}(C_{0u}-C_{0t})(\bar{C}_{0u}-\bar{C}_{0t})\Bigr]\equiv \frac{B \mathcal{\tilde{H}}_{s}}{\frac{\alpha' s}{4}}.\label{54}
\end{eqnarray}
\par Although this expression is not identical to Eqs.(49), (50) in the $s$-channel, if we compute the value of $\mathcal{H}_{s}$ 
at the pole $s=0$, its residue reads $\mathcal{H}_{s}|_{s=0}=-\mathcal{\tilde{H}}_{s}$. This implies that 
$\mathcal{\tilde{G}}_{4s}$ and $\mathcal{G}_{4s}$ are equivalent, since they have the same residue at the massless pole. The 
four-point function at the $s$-channel is consistent with the factorization property. The same is truth in the $t-$ and $u-$channel.
\subsection{Chiral Bosonic String}      
\subsubsection{Massless sector}
The product of two three-point amplitudes involving three massless states proceeds similarly as our discussion 
above of the type $II$ superstring. The intermediate state is a massless state and the modification with respect to the type $II$ 
superstring amplitude is very simple. In the $s$-channel we can write
\begin{eqnarray}
&& \tilde{G}_{40s} =  A \int \frac{d^{26}k}{(2 \pi)^{26}} \frac{[(2 \pi)^{26}]^{2}\delta^{26}(-k+k_{1}+k_{2})\delta^{26}(k+k_{3}+k_{4})}{-k^{2}+i \epsilon} \nonumber\\
&& \sum_{polarizations} A_{03}(k_{1},e_{1};k_{2},e_{2}; -k, e) A_{03}(k,e;k_{3},e_{3};k_{4},e_{4}).\label{55}
\end{eqnarray}
\par We integrate over $k$ to get the overall 26-dimensional delta function of momentum conservation along with the propagador 
replaced by the Mandelstam variable $s$. Using the projector just as before 
($\sum_{e} e_{\lambda_{1}} e_{\lambda_{2}}= \sum_{\tilde{e}} \tilde{e}_{\lambda_{1}} \tilde{e}_{\lambda_{2}} = \eta_{\lambda_{1} \lambda_{2}}$), we find
\begin{eqnarray}
&& \tilde{G}_{40s}= \frac{B}{\frac{\alpha' s}{4}} \Bigl[\bigl(\frac{\alpha' u}{4}\bigr)^{2} P_{s} 
\bar{P}_{s} 
+ \frac{\alpha'}{2} \bigl(\frac{\alpha' u}{4}\bigr) [P_{s}(\bar{C}_{u}-\bar{C}_{t}) + \bar{P}_{s}(C_{u}-C_{t})]\nonumber\\
&& + \bigl(\frac{\alpha'}{2}\bigr)^{2}(C_{u}-C_{t})(\bar{C}_{u}-\bar{C}_{t})\Bigr]\equiv \frac{B \tilde{H}_{0s}}{\frac{\alpha' s}{4}}.\label{56}
\end{eqnarray}
\par This form of the scattering amplitude does not coincide with that from Eqs. (\ref{27}), (\ref{28}). If Eq. (\ref{28}) is evaluated at $s=0$, its residue 
at the pole yields $H_{0s}|_{s=0}=-\tilde{H}_{0s}$. Thus, its residue relates the product of the two three-point 
amplitudes with the four massless scattering obtained from $KLT$ method, since 
$\tilde{G}_{40s} = \frac{B \tilde{H}_{0s}|_{s=0}}{\frac{\alpha' s}{4}}= (-1)\frac{B H_{0s}|_{s=0}}{\frac{\alpha' s}{4}}=G_{40s}$. The amplitude 
obtained using $KLT$ is more general, but restricting the attention to the residues at the pole of the numerator always relates them with the 
product of two three-point scattering amplitude. Of course, regular terms in $\tilde{G}_{40s}$ can always be neglected.
\subsubsection{Massive sector}
\par Let us analyze now the situation when a massive state in the three-point function is the intermediate state in the 
construction of the four-point amplitude. The symmetry between the right-mover and left-mover amplitudes 
relating tachyons and tardyons simplifies our discussion. 
\par The product of two three point amplitudes consisting of two massless states and one tachyon or tardyon as intermediate state can be 
treated in a unified fashion in close analogy with the discussion in the gravitational sector. Now, we have to perform the summation over the 
polarization of the internal tachyon or tardyon. To be specific let us start with the tachyon. It is important to remember 
that we shall use the amplitude $A_{021-}$ to build up the four massless scattering amplitude. Let this amplitude be denoted by 
$\tilde{G}_{4s-}$ in contrast with the definition $G_{4s-} = (-1)\frac{B H_{-s}}{1+\frac{\alpha' s}{4}}$ obtained 
from $KLT$ representation. 
\par The amplitude $\tilde{G}_{4s-}$ is defined by
\begin{eqnarray}
&& \tilde{G}_{4s-} = (-1) A \int \frac{d^{26}k}{(2 \pi)^{26}} \frac{[(2 \pi)^{26}]^{2}\delta^{26}(-k+k_{1}+k_{2})\delta^{26}(k+k_{3}+k_{4})}{-k^{2}+ \frac{4}{\alpha'} + i \epsilon} \nonumber\\
&& \sum_{polarizations} A_{021-}(k_{1},e_{1};k_{2},e_{2}; -k, \bar{\epsilon}) A_{021-}(k,\bar{\epsilon};k_{3},e_{3};k_{4},e_{4}).\label{57}
\end{eqnarray}
\par As already noted previously, differently from the case for the gravitational sector, the tachyon propagator is defined with a $(-1)$ sign in the above amplitude due to its ghost-like nature. We shall employ this convention in the tardyon sector as well later. The integration over 
the tachyon momentum $k$ yields similar terms as those obtained from the product of two three 
massless amplitudes, namely, the product of overall delta function, constants and a propagator $\frac{1}{1+\frac{\alpha' s}{4}}$. 
\par The fusion of the two three-point functions as a four-point amplitude can be given a meaning as follows. First, decompose the tensor 
polarization of the tachyon in terms of vector polarizations through 
$\bar{\epsilon}_{\mu_{1} \nu_{1}}=\bar{\epsilon}_{\mu_{1}} \bar{\epsilon}_{\nu_{1}}$. Second, define the projector as the 
contraction of two vector polarizations belonging to distinct three-point amplitudes as 
$\sum_{\bar{\epsilon}} \bar{\epsilon}_{\mu_{1}} \bar{\epsilon}_{\mu_{2}} = \Pi_{\mu_{1} \mu_{2}} \equiv \eta_{\mu_{1} \mu_{2}} 
-\frac{k_{\mu_{1}} k_{\mu_{2}}}{k^{2}}$. This has the advantage of keeping the transversality (gauge invariance) of the tachyon on-shell, since 
$k^{\mu_{1}} \sum_{\bar{\epsilon}} \bar{\epsilon}_{\mu_{1}} \bar{\epsilon}_{\mu_{2}}=0$. From now on, we adopt the projector computed on-shell. For 
the tachyon, it is given by $\Pi_{\mu_{1} \mu_{2}} = \eta_{\mu_{1} \mu_{2}} 
+\frac{\alpha'}{4}k_{\mu_{1}} k_{\mu_{2}}$ (whereas for the tardyon it is given by  $\Pi_{\mu_{1} \mu_{2}} = \eta_{\mu_{1} \mu_{2}} 
-\frac{\alpha'}{4}k_{\mu_{1}} k_{\mu_{2}}$). Since there is one tensor 
polarization in each three-point amplitude, the question is how to melt them together inspired by a similar situation  taking place in massive gravity theories. Indeed, a direct comparison with standard massive gravity theories \cite{KH} can indicate the trend to follow henceforth. The classical equations of motion for the massive gravity derived from a 
field-theoretic action, imply Eqs. (2.11) from Ref. \cite{KH}  
for our would-be massive graviton $E_{\mu \nu}$ ($(h_{\mu \nu})_{his}= (E_{\mu \nu})_{ours}$) , namely
\begin{subequations}
\begin{eqnarray}
&& (\partial^{\rho}\partial_{\rho} - m^{2}) E_{\mu \nu}=0,\\
&&  \partial^{\mu} E_{\mu \nu}=0,\\
&& E= E^{\mu}_{\mu}=0.
\end{eqnarray}
\end{subequations}
After quantization the completeness relations are given (according to our conventions, utilizing the normalization defined in Ref \cite{KH}) by the expression
\begin{eqnarray} 
&& \sum_{E} E_{\mu_{1} \nu_{1}} E_{\mu_{2} \nu_{2}} = \frac{1}{2}\bigl(\eta_{\mu_{1} \mu_{2}} + 
\frac{k_{\mu_{1}} k_{\mu_{2}}}{m^{2}}\bigr) \bigl(\eta_{\nu_{1} \nu_{2}} +\frac{k_{\nu_{1}} k_{\nu_{2}}}{m^{2}}\bigr) 
+ \frac{1}{2}\bigl(\eta_{\mu_{1} \nu_{2}} + \frac{k_{\mu_{1}} k_{\nu_{2}}}{m^{2}}\bigr) \bigl(\eta_{\nu_{1} \mu_{2}} \nonumber\\
&&+ \frac{k_{\nu_{1}} k_{\mu_{2}}}{m^{2}}\bigr) -\frac{1}{D-1} \bigl(\eta_{\mu_{1} \nu_{1}} +\frac{k_{\mu_{1}} k_{\nu_{1}}}{m^{2}}\bigr) \bigl(\eta_{\mu_{2} \nu_{2}} +\frac{k_{\mu_{2}} k_{\nu_{2}}}{m^{2}}\bigr).\label{63} 
\end{eqnarray}
\par Our problem is a bit different: the tachyon and the tardyon are indeed massive spin-2 fields, although with different signs for the squared mass. Their classical equations of motion are identical to that above for the massive tensor field $E_{\mu \nu}$ except for the trace, since so far we have no information about it. One alternative is try to find the trace for the tachyon and the tardyon by postulating a completeness relation to them similar to that for the massive field  $E_{\mu \nu}$, but replacing the masses of the tachyon and tardyon to their values on-shell.   
\par Inspired in the massive gravity and its similarity with the spin-2 massive particles in the spectrum of the chiral bosonic string, we now make use of the projector/completeness relation (with a different normalization) 
\begin{eqnarray} 
&& \sum_{\bar{\epsilon}} \bar{\epsilon}_{\mu_{1} \nu_{1}} \bar{\epsilon}_{\mu_{2} \nu_{2}} = \bigl(\eta_{\mu_{1} \mu_{2}} -\frac{\alpha'}{4} k_{\mu_{1}} k_{\mu_{2}}\bigr) \bigl(\eta_{\nu_{1} \nu_{2}} -\frac{\alpha'}{4} k_{\nu_{1}} k_{\nu_{2}}\bigr) + \bigl(\eta_{\mu_{1} \nu_{2}} -\frac{\alpha'}{4} k_{\mu_{1}} k_{\nu_{2}}\bigr) \bigl(\eta_{\nu_{1} \mu_{2}} 
-\frac{\alpha'}{4} k_{\nu_{1}}
\times\nonumber\\
&& k_{\mu_{2}}\bigr) - \tilde{a} \bigl(\eta_{\mu_{1} \nu_{1}} -\frac{\alpha'}{4} k_{\mu_{1}} k_{\nu_{1}}\bigr) 
\bigl(\eta_{\mu_{2} \nu_{2}} -\frac{\alpha'}{4} k_{\mu_{2}} k_{\nu_{2}}\bigr), \label{58}
\end{eqnarray}
for the tachyon polarizations. If the construction is consistent, we should obtain the value of the trace for tachyon 
and tardyon in the end of the calculation. In other words, the constant $\tilde{a}$ can be obtained in the end when comparing the result using this recipe to get the 
four-point amplitude with the one coming from the $KLT$ representation. In the computation of the amplitude $\tilde{G}_{4s-}$, we 
neglect all terms proportional to $(1+\frac{\alpha' s}{4})$, since they are regular and do not contribute to the pole of the propagator.
\par An efficient way to work with the projector just defined is to isolate the different powers of $\alpha'$ in it, before actually applying it 
in the expression of the amplitude. Utilizing the set of steps above and after organizing the several terms, the amplitude can be 
written in the form:
\begin{eqnarray}
&& \tilde{G}_{4s-} = (-1) B \frac{[H_{-s} -  P_{s}(\frac{\alpha'}{2}\tilde{E}_{1}+\tilde{E}_{2}(\tilde{a}))]}{1+\frac{\alpha' s}{4}},\label{59}\\
&& \tilde{E}_{1} = \frac{\alpha' s}{4} \Bigl[\bigl(\tilde{e}_{1}.\tilde{e}_{2} 
+ \frac{\alpha'}{2} k_{2}.\tilde{e}_{1} k_{1}.\tilde{e}_{2}\bigr)k_{4}.\tilde{e}_{3} k_{3}.\tilde{e}_{4} + \bigl(\tilde{e}_{3}.\tilde{e}_{4} 
+ \frac{\alpha'}{2} k_{4}.\tilde{e}_{3} k_{3}.\tilde{e}_{4}\bigr)k_{2}.\tilde{e}_{1} k_{1}.\tilde{e}_{2} - \frac{\alpha'}{2}(k_{4}.\tilde{e}_{1} \nonumber\\
&& k_{3}.\tilde{e}_{2} k_{1}.\tilde{e}_{3} k_{2}.\tilde{e}_{4} + k_{3}.\tilde{e}_{1} k_{4}.\tilde{e}_{2} k_{2}.\tilde{e}_{3} k_{1}.\tilde{e}_{4}) - \frac{1}{2}\bigl(\tilde{e}_{1}.\tilde{e}_{2} + \frac{\alpha'}{2} k_{2}.\tilde{e}_{1} k_{1}.\tilde{e}_{2} \bigr)(k_{1}.\tilde{e}_{3} k_{1}.\tilde{e}_{4}+ k_{2}.\tilde{e}_{3} k_{2}.\tilde{e}_{4})\nonumber\\
&&  - \frac{1}{2}\bigl(\tilde{e}_{3}.\tilde{e}_{4} + \frac{\alpha'}{2} k_{4}.\tilde{e}_{3} k_{3}.\tilde{e}_{4} \bigr)(k_{3}.\tilde{e}_{1} k_{3}.\tilde{e}_{2}+ k_{4}.\tilde{e}_{1} k_{4}.\tilde{e}_{2})\Bigr] + \frac{1}{2} \bigl(\tilde{e}_{1}.\tilde{e}_{2} + \frac{\alpha'}{2} k_{2}.\tilde{e}_{1} k_{1}.\tilde{e}_{2} \bigr)(k_{2}.\tilde{e}_{3} k_{1}.\tilde{e}_{4}\nonumber\\
&& + k_{1}.\tilde{e}_{3} k_{2}.\tilde{e}_{4}) + \frac{1}{2}\bigl(\tilde{e}_{3}.\tilde{e}_{4} + \frac{\alpha'}{2} k_{4}.\tilde{e}_{3} k_{3}.\tilde{e}_{4} \bigr)(k_{4}.\tilde{e}_{1} k_{3}.\tilde{e}_{2}+ k_{3}.\tilde{e}_{1} k_{4}.\tilde{e}_{2}) +\frac{\alpha'}{2}(k_{4}.\tilde{e}_{1} k_{1}.\tilde{e}_{2} k_{2}.\tilde{e}_{3} k_{3}.\tilde{e}_{4}\nonumber\\
&&  + k_{3}.\tilde{e}_{1} k_{1}.\tilde{e}_{2} k_{4}.\tilde{e}_{3} k_{2}.\tilde{e}_{4} 
+ k_{2}.\tilde{e}_{1} k_{4}.\tilde{e}_{2} k_{1}.\tilde{e}_{3} k_{3}.\tilde{e}_{4} 
+ k_{2}.\tilde{e}_{1} k_{3}.\tilde{e}_{2} k_{4}.\tilde{e}_{3} k_{1}.\tilde{e}_{4}
- k_{3}.\tilde{e}_{1} k_{4}.\tilde{e}_{2} k_{1}.\tilde{e}_{3} k_{2}.\tilde{e}_{4}\nonumber\\
&& - k_{4}.\tilde{e}_{1} k_{3}.\tilde{e}_{2} k_{2}.\tilde{e}_{3} k_{1}.\tilde{e}_{4}),\label{60}
\end{eqnarray}
\par Before defining $\tilde{E}_{2}(\tilde{a})$, let us pause to make a comparison with the gravitational sector. The factor $H_{-s}$ is 
rigorously the same as the one appearing in $G_{4s-} \equiv (-1) \frac{B H_{-s}}{1+\frac{\alpha' s}{4}}$. The role played by the 
residue at the tachyon pole here is different: all we have to show is that the extra terms in the above expression for 
$ \tilde{G}_{4s-}$ cancel at the tachyon pole. This fixes the value of the constant $\tilde{a}$, which is, as explained before the indirect determination of the trace of the state described by the tachyon vertex operator.  
\par The object $\tilde{E}_{2}(\tilde{a})$ is defined by:
\begin{eqnarray}
&& \tilde{E}_{2}(\tilde{a}) = -\tilde{a}(\tilde{e}_{1}.\tilde{e}_{2} -\alpha' k_{2}.\tilde{e}_{1} k_{1}.\tilde{e}_{2})(\tilde{e}_{3}.\tilde{e}_{4} -\alpha' k_{4}.\tilde{e}_{3} k_{3}.\tilde{e}_{4}) + \frac{\tilde{a}}{2}\bigl(\frac{\alpha' s}{4}\bigr)\Bigl[(\tilde{e}_{1}.\tilde{e}_{2} -\alpha' k_{2}.\tilde{e}_{1} k_{1}.\tilde{e}_{2})\times \nonumber\\
&& \bigl(\tilde{e}_{3}.\tilde{e}_{4} +\frac{\alpha'}{2} k_{4}.\tilde{e}_{3} k_{3}.\tilde{e}_{4}\bigr) + (\tilde{e}_{3}.\tilde{e}_{4} 
-\alpha' k_{4}.\tilde{e}_{3} k_{3}.\tilde{e}_{4})\bigl(\tilde{e}_{1}.\tilde{e}_{2} +\frac{\alpha'}{2} k_{2}.\tilde{e}_{1} k_{1}.\tilde{e}_{2}\bigr)\Bigr] + \frac{\tilde{a}}{4}\Bigl[\tilde{e}_{1}.\tilde{e}_{2}\bigl(\tilde{e}_{3}.\tilde{e}_{4} +\frac{\alpha'}{2} \nonumber\\
&& \times k_{4}.\tilde{e}_{3} k_{3}.\tilde{e}_{4}\bigr) + \tilde{e}_{3}.\tilde{e}_{4} \bigl(\tilde{e}_{1}.\tilde{e}_{2} +\frac{\alpha'}{2} k_{2}.\tilde{e}_{1} k_{1}.\tilde{e}_{2}\bigr)\Bigr]
-\frac{\tilde{a} \alpha'}{4}[(\tilde{e}_{1}.\tilde{e}_{2} +\alpha' k_{2}.\tilde{e}_{1} k_{1}.\tilde{e}_{2})k_{4}.\tilde{e}_{3} k_{3}.\tilde{e}_{4}
+(\tilde{e}_{3}.\tilde{e}_{4} \nonumber\\
&& +\alpha' k_{4}.\tilde{e}_{3} k_{3}.\tilde{e}_{4}) k_{2}.\tilde{e}_{1} k_{1}.\tilde{e}_{2}] -\frac{\alpha'}{4}\Bigl(1+\bigl(\frac{\tilde{a} \alpha' s}{8}\bigr) - (1-\tilde{a}) \Bigl(1-\bigl(\frac{\alpha' s}{4}\bigr)\Bigr)\Bigr)\Bigl[\bigl(\tilde{e}_{1}.\tilde{e}_{2} + \frac{\alpha'}{2} k_{2}.\tilde{e}_{1} k_{1}.\tilde{e}_{2}\bigr)\nonumber\\
&& \times k_{4}.\tilde{e}_{3} k_{3}.\tilde{e}_{4}
+\bigl(\tilde{e}_{3}.\tilde{e}_{4} + \frac{\alpha'}{2} k_{4}.\tilde{e}_{3} k_{3}.\tilde{e}_{4}\bigr) k_{2}.\tilde{e}_{1} k_{1}.\tilde{e}_{2}\Bigr] + \alpha^{' 2}\Bigl[\tilde{a}+ \bigl(\frac{\alpha' s}{16}\bigr)\Bigr]k_{2}.\tilde{e}_{1} k_{1}.\tilde{e}_{2} k_{4}.\tilde{e}_{3} k_{3}.\tilde{e}_{4} \nonumber\\
&& + \Bigl(\frac{2-\tilde{a}}{16}\Bigr)\Bigl(\tilde{e}_{1}.\tilde{e}_{2} -\frac{\alpha'}{2} k_{2}.\tilde{e}_{1} k_{1}.\tilde{e}_{2}\Bigr)
\Bigl(\tilde{e}_{3}.\tilde{e}_{4} -\frac{\alpha'}{2} k_{4}.\tilde{e}_{3} k_{3}.\tilde{e}_{4}\Bigr).\label{61}
\end{eqnarray}
\par These extra terms become quite simple at the tachyon pole $\frac{\alpha' s}{4}=-1$. Indeed, when we compute the residue of $\tilde{E}_{1}$ 
and $\tilde{E}_{2}$ at the tachyon pole, we find that $[\frac{\alpha'}{2}\tilde{E}_{1}+\tilde{E}_{2}(\tilde{a})]_{\frac{\alpha' s}{4}=-1}=0$ for 
$\tilde{a}=\frac{2}{25}$ (or $\tilde{a}=\frac{2}{D-1}$ when $D=26$). Therefore $G_{4s-}$ from $KLT$ method is equivalent to the amplitude 
$\tilde{G}_{4s-}$ calculated from the product of two three-point scattering amplitudes, {\it provided the tachyon polarization is traceless}.
\par Let us follow the same steps using analogous arguments when the tardyon is the intermediate state. The four-point amplitude 
$\tilde{G}_{4s+}$ obtained from the product of two three-point functions now reads 
\begin{eqnarray}
&& \tilde{G}_{4s+} = (-1) A \int \frac{d^{26}k}{(2 \pi)^{26}} \frac{[(2 \pi)^{26}]^{2}\delta^{26}(-k+k_{1}+k_{2})\delta^{26}(k+k_{3}+k_{4})}{-k^{2}- \frac{4}{\alpha'} + i \epsilon} \nonumber\\
&& \sum_{polarizations} A_{021+}(k_{1},e_{1};k_{2},e_{2}; -k, \bar{\epsilon}) A_{021+}(k,\bar{\epsilon};k_{3},e_{3};k_{4},e_{4}).\label{62}
\end{eqnarray}
\par Perform the integral over $k$ just as before. Now use the following completeness relation for the tardyon polarizations 
$\epsilon_{\mu \nu}$
\begin{eqnarray} 
&& \sum_{\epsilon} \epsilon_{\mu_{1} \nu_{1}} \epsilon_{\mu_{2} \nu_{2}} = \bigl(\eta_{\mu_{1} \mu_{2}} +\frac{\alpha'}{4} k_{\mu_{1}} k_{\mu_{2}}\bigr) \bigl(\eta_{\nu_{1} \nu_{2}} +\frac{\alpha'}{4} k_{\nu_{1}} k_{\nu_{2}}\bigr) + \bigl(\eta_{\mu_{1} \nu_{2}} +\frac{\alpha'}{4} k_{\mu_{1}} k_{\nu_{2}}\bigr) \bigl(\eta_{\nu_{1} \mu_{2}} +\frac{\alpha'}{4} k_{\nu_{1}}\times\nonumber\\
&& k_{\mu_{2}}\bigr) - a \bigl(\eta_{\mu_{1} \nu_{1}} +\frac{\alpha'}{4} k_{\mu_{1}} k_{\nu_{1}}\bigr) \bigl(\eta_{\mu_{2} \nu_{2}} +\frac{\alpha'}{4} k_{\mu_{2}} k_{\nu_{2}}\bigr).\label{63}
\end{eqnarray}
\par Proceeding in the same way as discussed for the tachyon, and discarding regular terms at the tardyon pole (proportional to 
$(1-\frac{\alpha's}{4}$)) we find the following value for the scattering amplitude 
 \begin{eqnarray}
&& \tilde{G}_{4s+} = (-1) B \frac{[H_{+s} - \bar{P}_{s}(\frac{\alpha'}{2} E_{1}+ E_{2}(a))]}{1-\frac{\alpha' s}{4}},\label{64}
\end{eqnarray}
where
\begin{eqnarray}
&& E_{1} = \frac{\alpha' s}{4} \Bigl[\bigl(e_{1}.e_{2} 
- \frac{\alpha'}{2} k_{2}.e_{1} k_{1}.e_{2}\bigr)k_{4}.e_{3} k_{3}.e_{4} + \bigl(e_{3}.e_{4} 
- \frac{\alpha'}{2} k_{4}.e_{3} k_{3}.e_{4}\bigr)k_{2}.e_{1} k_{1}.e_{2} + \frac{\alpha'}{2}(k_{4}.e_{1} \nonumber\\
&& k_{3}.e_{2} k_{1}.e_{3} k_{2}.e_{4} + k_{3}.e_{1} k_{4}.e_{2} k_{2}.e_{3} k_{1}.e_{4}) - \frac{1}{2}\bigl(e_{1}.e_{2} 
- \frac{\alpha'}{2} k_{2}.e_{1} k_{1}.e_{2} \bigr)(k_{1}.e_{3} k_{1}.e_{4}+ k_{2}.e_{3} k_{2}.e_{4})\nonumber\\
&&  - \frac{1}{2}\bigl(e_{3}.e_{4} - \frac{\alpha'}{2} k_{4}.e_{3} k_{3}.e_{4} \bigr)(k_{3}.e_{1} k_{3}.e_{2}+ k_{4}.e_{1} k_{4}.e_{2})\Bigr] - \frac{1}{2} \bigl(e_{1}.e_{2} - \frac{\alpha'}{2} k_{2}.e_{1} k_{1}.e_{2} \bigr)(k_{2}.e_{3} k_{1}.e_{4}\nonumber\\
&& + k_{1}.e_{3} k_{2}.e_{4}) - \frac{1}{2}\bigl(e_{3}.e_{4} - \frac{\alpha'}{2} k_{4}.e_{3} k_{3}.e_{4} \bigr)(k_{4}.e_{1} k_{3}.e_{2}+ k_{3}.e_{1} 
k_{4}.e_{2}) +\frac{\alpha'}{2}(k_{4}.e_{1} k_{1}.e_{2} k_{2}.e_{3} k_{3}.e_{4}\nonumber\\
&&  + k_{3}.e_{1} k_{1}.e_{2} k_{4}.e_{3} k_{2}.e_{4} 
+ k_{2}.e_{1} k_{4}.e_{2} k_{1}.e_{3} k_{3}.e_{4} 
+ k_{2}.e_{1} k_{3}.e_{2} k_{4}.e_{3} k_{1}.e_{4}
- k_{3}.e_{1} k_{4}.e_{2} k_{1}.e_{3} k_{2}.e_{4}\nonumber\\
&& - k_{4}.e_{1} k_{3}.e_{2} k_{2}.e_{3} k_{1}.e_{4}),\\
&& E_{2}(a) = - a(e_{1}.e_{2} +\alpha' k_{2}.e_{1} k_{1}.e_{2})(e_{3}.e_{4} +\alpha' k_{4}.e_{3} k_{3}.e_{4}) - \frac{a}{2}\bigl(\frac{\alpha' s}{4}\bigr)\Bigl[(e_{1}.e_{2} +\alpha' k_{2}.e_{1} k_{1}.e_{2})\times \nonumber\\
&& \bigl(e_{3}.e_{4} -\frac{\alpha'}{2} k_{4}.e_{3} k_{3}.e_{4}\bigr) + (e_{3}.e_{4} 
+\alpha' k_{4}.e_{3} k_{3}.e_{4})\bigl(e_{1}.e_{2} -\frac{\alpha'}{2} k_{2}.e_{1} k_{1}.e_{2}\bigr)\Bigr] 
+ \frac{a}{4}\Bigl[e_{1}.e_{2}\bigl(e_{3}.e_{4} -\frac{\alpha'}{2} \nonumber\\
&& \times k_{4}.e_{3} k_{3}.e_{4}\bigr) + e_{3}.e_{4} \bigl(e_{1}.e_{2} -\frac{\alpha'}{2} k_{2}.e_{1} k_{1}.e_{2}\bigr)\Bigr]
+\frac{a \alpha'}{4}[(e_{1}.e_{2} -\alpha' k_{2}.e_{1} k_{1}.e_{2})k_{4}.e_{3} k_{3}.e_{4}
+(e_{3}.e_{4} \nonumber\\
&& -\alpha' k_{4}.e_{3} k_{3}.e_{4}) k_{2}.e_{1} k_{1}.e_{2}] +\frac{\alpha'}{4}\Bigl(1-\bigl(\frac{a \alpha' s}{8}\bigr) - (1-a) \Bigl(1+\bigl(\frac{\alpha' s}{4}\bigr)\Bigr)\Bigr)\Bigl[\bigl(e_{1}.e_{2} - \frac{\alpha'}{2} k_{2}.e_{1} k_{1}.e_{2}\bigr)\nonumber\\
&& \times k_{4}.e_{3} k_{3}.e_{4}
+\bigl(e_{3}.e_{4} - \frac{\alpha'}{2} k_{4}.e_{3} k_{3}.e_{4}\bigr) k_{2}.e_{1} k_{1}.e_{2}\Bigr] + \alpha^{' 2}\Bigl[a- \bigl(\frac{\alpha' s}{16}\bigr)\Bigr]k_{2}.e_{1} k_{1}.e_{2} k_{4}.e_{3} k_{3}.e_{4} \nonumber\\
&& + \Bigl(\frac{2-a}{16}\Bigr)\Bigl(e_{1}.e_{2} +\frac{\alpha'}{2} k_{2}.e_{1} k_{1}.e_{2}\Bigr)
\Bigl(e_{3}.e_{4} +\frac{\alpha'}{2} k_{4}.e_{3} k_{3}.e_{4}\Bigr).\label{66}
\end{eqnarray}
\par At the tardyon pole, $[\frac{\alpha'}{2} E_{1}+ E_{2}(a)]_{\frac{\alpha' s}{4}=1}=0$ for $a=\frac{2}{D-1}$ with $D=26$. At this value of $a$, which is the 
same value of $\tilde{a}$ in the tachyon pole, the scattering amplitude obtained via $KLT$ is identical to that computed using factorization 
arguments as discussed above. These conditions for the constants simply imply that the tachyon and tardyon polarizations are traceless. This is a non-trivial fact, since differently from the massive gravity presented in Ref. \cite{KH}, there was no classical equation of motion determining the trace to begin with. Rather, here it is determined through a quantum 
consistency condition involving scattering amplitudes evaluated in two different manners. In particular, the critical dimension where the original string "lives" (along with the massive vertex operators stemming from its spectrum) does not imply that the field theory originating from this string is only consistent in this critical dimension. But only at this critical dimension the factorization property is satisfied.  
\section{Scattering amplitude for chiral heterotic superstrings at the massless sector}
\par A simple application of the method developed above for bosonic and (supersymmetric) type $II$ strings can be outlined for heterotic 
superstrings using just the product of two three-point amplitudes. We focus only on the massless sector, but the discussion can also be 
extended to compute the scattering amplitude corresponding to four massless states either in the tachyonic or the tardyonic sector.
\par The heterotic superstring resembles the bosonic string in that either the tachyon or the tardyon is present in the spectrum in the 
left- or right-mover sector. With either choice, the other sector is similar to the type $II$ superstring. In both string 
specimes, we have just shown the equivalence between the product of the 2 three-point amplitudes and four-point scattering using $KLT$ 
representation. This permits us to construct the scattering amplitude of four massless states very easily as follows.  
\par Restricting our attention to the heterotic superstring with the tardyon as its only massive state, we now evaluate the 
product of 2 three-point heterotic superstring amplitudes with a massless state as intermediate state. It can be shown that the amplitude, 
in the $s$ channel for instance, reads
\begin{eqnarray}
&& \tilde{\mathcal{G}}_{4h}= \frac{B}{\frac{\alpha's}{4}}\Bigl[\Bigl(\frac{\alpha'}{4}\Bigr)^{2} u^{2} P_{s} \bar{\mathcal{P}}_{s} 
- \frac{\alpha'}{2}\Bigl[\frac{\alpha' u}{4}(P_{s} \bar{C}_{0t} + \bar{\mathcal{P}}_{s} C_{t})-\frac{\alpha' u}{4}(P_{s} \bar{C}_{0u} \nonumber\\
&& + \bar{\mathcal{P}}_{s} C_{u})\Bigr] + \Bigl(\frac{\alpha'}{2}\Bigr)^{2}(C_{u} - C_{t})(\bar{C}_{0u} -\bar{C}_{0t})\Bigr].\label{67}
\end{eqnarray}
\par Using $s+t+u=0$ at the pole $s=0$ we find that the last expression can be identified with the scattering amplitude of four heterotic 
massless states with a massless as intermediate state in the form 
\begin{eqnarray}
&& \tilde{\mathcal{G}}_{4h}= (-1)\frac{B}{\frac{\alpha's}{4}}\Biggl[(\frac{\alpha'}{4})^{2} tu P_{s} \bar{\mathcal{P}}_{s} 
- \frac{\alpha'}{2}\Bigl(\frac{\alpha' t}{4}\Bigl(P_{s} \bar{C}_{0t} + \bar{\mathcal{P}}_{s} C_{t}\Bigl)
+ \frac{\alpha' u}{4}\Bigl(P_{s} \bar{C}_{0u} + \bar{\mathcal{P}}_{s} C_{u}\Bigr)\Bigr) \nonumber\\
&& - \frac{\alpha^{' 2}}{4}(C_{u} - C_{t})(\bar{C}_{0u} - \bar{C}_{0t})\Biggr].\label{68}
\end{eqnarray}
Of course, this result can be easily obtained using the $KLT$ method. 
\section{Conclusion}
\par In this work we computed the scattering amplitudes associated with four massless external states for the closed chiral string in its 
bosonic, type $II$ and heterotic versions. The bosonic and type $II$ superstrings were obtained using the $KLT$ method. The results were 
checked using factorization arguments: in each channel, the product of two three-point scattering amplitudes agrees with the $KLT$ method. When the 
intermediate state is massless, the amplitude obtained using these two techniques agrees with each other trivially. 
\par When the intermediate state is massive, the two results are equivalent provided the polarization of the spin-2 internal state 
(either tachyon or tardyon) is traceless. In particular, the factorization argument regarding the tracelessness of the massive intermediate state 
leads directly to the critical dimension of the bosonic string. Previous arguments either used particular values of the intercept of the Regge 
trajectories as originally proposed in Ref. \cite{Virasoro} (see also \cite{Veneziano}) along with the Virasoro algebra in the elimination of 
negative norm states \cite{DDF,Brower,GT} or employed the solution of the integral corresponding to the four-point function in terms of the Euler 
Beta function \cite{Frampton,Veneziano}. To the best of our knowledge, getting the value of the critical dimension 
without making explicit reference to the Virasoro algebra (or Lorentz invariance in particular gauges) is a new feature stemming from the chiral 
bosonic string. 
\par This feature is possible because the dimension of spacetime appears in the polarization sum for 
massive traceless spin-2 fields in massive gravity, since the traceless condition arises from the 
classical equations of motion \cite{KH}. Here, we did not start from a classical equation of motion for the trace, 
but obtained the traceless condition from the identity between $KLT$ method in getting the 4-point amplitude with the factorization of two 3-point amplitudes with massive particles as intermediate states. After the identification of the coefficient of the trace with the standard outcome from massive gravity we conclude that $D=D_{c}=26$. Though the method presented here could be applied to normal bosonic strings using the factorization of two 3-point amplitudes with an intermediate massive spin-2 state, the comparison with the 4-point obtained from $KLT$ might not be as simple as here, since in that case the infinite number of intermediate states in the scattering amplitude of four massless states could 
make the task (if not impossible) very difficult. Further investigation on this point might offer a definite answer to this consideration. All chiral string theories treated in the present paper have scattering amplitudes manifestly gauge invariant. 
\par We discussed solely the massless sector of the heterotic string but the massive sector corresponding to a massive 
intermediate state can be approached with the results derived in the present work.
\par We have shown in fact that tachyon and tardyon are spin-$2$ ghosts of the bosonic string. Note that this just agrees with the criteria 
from Refs. \cite{HSZ,HNZ} and string field theory. (Indeed, Ref. \cite{HNZ} presents extra massive scalar states as well that do not take place in 
the spectra of the work discussed herein.) Contrarily, the schematic 
$KLT$ analysis without explicit separation of 
all simple poles from Ref. \cite{S2} concludes that the massive tardyon is not a ghost. They used a fixed value of one of 
the Mandelstam variable characteristic of a given channel in the prefactor $\frac{1}{stu}$ corresponding to a certain 
massive state and obtained a double pole involving the massless and the other massive state in another channel. This was not 
followed by the disentanglement of the resulting double pole in terms of simple poles of the massless and massive states 
along with the full development of its coefficient. The resulting analysis is confusing. Our 
treatment here is more streamlined: the given state is analyzed in a given channel and its residue (involving only single 
poles) is computed in the same channel. Furthermore, in the present paper, we shared the vision that $KLT$ alone is not 
enough to make a definitive statement about the ghost character of the tardyon. Rather, we witness this character through 
the product of two three-point scattering amplitudes computed at the residue of the massive tardyon pole in the appropriate 
channel. The incorrect conclusion from Ref. \cite{S2} can thus be traced back not only to the combination of double poles in 
a different channel from that original one was trying to describe in the first place but also to the odd claim for the 
chiral bosonic string that $KLT$ itself suffices to deduce the ghost-like feature of the tardyon.  
\par Although the intermediate steps in the computation of the amplitude are involved and require much toil, the simplicity 
of the answers resembling those outcomes from field theory makes the study of further properties of chiral strings 
worthwhile. A desirable property that is 
not explicit in the amplitude just studied is that the scattering equations are not manifest in the output for the scattering amplitudes. 
Other gauge choices in the worldsheet fields produce these equations automatically and might be interesting to study, like in the case of 
left-handed strings \cite{S1}. It would be important to check whether the final form of the amplitudes have a simple field-theoretic resemblance as 
those studied in the present paper.
\par  Other aspects that can be pursued are the construction of low energy effective actions for the states of the spectrum of chiral strings 
\cite{HSZ,HNZ}. Since there are a finite number of states, an effective action from field theory can be constructed in principle in order to 
investigate how the states of the spectrum interact among each other. We leave these topics for future work.        

\section{Acknowledgements}
\par M.M.L. thanks CNPq grant number 232352/2014-3 for partial financial support. W.S. is supported in part by National Science Foundation 
Grant No. PHY-1316617. 

\appendix   
\section{Some properties of $C_{s}, \bar{C}_{s}, C_{t}, \bar{C}_{t}, C_{u}, \bar{C}_{u}$}
\par As pointed out in the main text, although the amplitudes in each sector apparently have a $stu$ symmetry in the different channels, 
the $O(\alpha')$ corrections only appear in $C_{1t}$ and $C_{1u}$. Hence the $stu$ symmetry is not manifest in the amplitudes of the tachyonic, 
gravitational and tardyonic sectors. Here we prove that all amplitudes computed at the residues of the appropriate poles at each channel are 
$stu$ symmetric.  
\par It is obvious that the quantities $C_{0s}, C_{0t}, C_{0u}$ (and $\bar{C}_{0s}, \bar{C}_{0t}, \bar{C}_{0u}$) are explicitly $stu$ 
symmetric. In order to see this, in the expressions defining these objects given in the main text, it is easy to verify that if we change 
the momentum and polarization labels according to the rule {\it i)} $1 \rightarrow 3$, the Mandelstam variables undergo the change 
$s \rightarrow t$ and $C_{0s}\rightarrow C_{0t}$ (and vice-versa); {\it ii)} $1 \rightarrow 4$, then $s \rightarrow u$ and 
$C_{0s}\rightarrow C_{0u}$ (and vice-versa); {\it iii)} $3 \rightarrow 4$, then $t \rightarrow u$ and 
$C_{0t}\rightarrow C_{0u}$ (and vice-versa).
\par The several amplitudes of the bosonic string are $stu$ symmetric, except for the terms involving  
$C_{1t}, \bar{C}_{1t}, C_{1u}, \bar{C}_{1u}$ for they do not have an analogue in the $s$ index. We focus our attention in the analysis of 
these terms and their combinations which are manifestly $stu$ symmetric. 
\par We commence by writing them down. We consider only the left-mover contributions. (The right-mover contributions have exactly 
the same momentum dependence and the right-mover polarizations replace the left-mover ones, so the argument carries out easily for them). They 
read
\begin{eqnarray}
&& C_{1t} (e) = k_{3}.e_{1} k_{4}. e_{2}(k_{1}.e_{3} k_{1}.e_{4} + k_{2}.e_{3} k_{2}.e_{4}) -  (k_{3}.e_{1} k_{1}.e_{2} + k_{2}.e_{1} k_{4}.e_{2}) k_{1}. e_{3}  k_{2}.e_{4} ,\label{A1}\\
 && C_{1u}(e)= (k_{3}.e_{1} k_{3}.e_{2} + k_{4}.e_{1} k_{4}.e_{2}) k_{2}.e_{3} k_{1}.e_{4} - k_{4}.e_{1} k_{3}.e_{2}(k_{4}.e_{3} k_{1}.e_{4} + k_{2}.e_{3} k_{3}.e_{4}).\label{A2}
\end{eqnarray}
\par Look at $C_{1u}$ and perform the change $3 \rightarrow 4$. Using momentum conservation and gauge invariance in the form $k_{i}.e_{i}=0$, 
it is easy to show that the above expression can be transformed into $C_{1t}$, 
\begin{eqnarray}
&& C_{1u}(3 \rightarrow 4) = (k_{4}.e_{1} k_{4}.e_{2} + k_{3}.e_{1} k_{3}.e_{2}) k_{2}.e_{4} k_{1}.e_{3} - k_{3}.e_{1} k_{4}.e_{2}(k_{3}.e_{4} k_{1}.e_{3} + k_{2}.e_{4} k_{4}.e_{3})\nonumber\\
&& = C_{1t}.\label{A3}
\end{eqnarray} 
Hence the above expression can be viewed as another manner of writing $C_{1t}$. By the same token, it can be easily shown that 
\begin{eqnarray}
&& C_{1t}(3 \rightarrow 4) = k_{4}.e_{1} k_{3}. e_{2}(k_{1}.e_{4} k_{1}.e_{3} + k_{2}.e_{4} k_{2}.e_{3}) -  (k_{4}.e_{1} k_{1}.e_{2} + k_{2}.e_{1} k_{3}.e_{2}) k_{1}. e_{4}  k_{2}.e_{3} \nonumber\\
&& = C_{1u} .\label{A4}
\end{eqnarray} 
\par Now consider the scattering amplitude of the massless sector in the $s$-channel Eq. (\ref{56}). As discussed in the bulk of the paper, this 
expression corresponds to $G_{40s}$ with numerator $H_{0s}$ computed at the residue of the pole at $s=0$. A potential problem in the violation 
of the $stu$ symmetry comes from the term(s) $C_{u}-C_{t}$ $(\bar{C}_{u}-\bar{C}_{t})$, specifically from the contributions $(C_{1u}-C_{1t})$, 
etc., by the aforementioned reasons.  Using the above expressions, we 
find that $(C_{1u}-C_{1t})(1 \rightarrow 3)= C_{1u}$. (As $C_{1s}=0$, we could also have written $(C_{1u}-C_{1t})(1 \rightarrow 3)=C_{1u}-C_{1s}$). 
Thus, performing the transposition $1 \rightarrow 3$ and using the facts highlighted above we find
\begin{eqnarray}
&& \tilde{G}_{40s}(1\rightarrow 3)= \frac{B}{\frac{\alpha' t}{4}} \Bigl[\bigl(\frac{\alpha' u}{4}\bigr)^{2} P_{t} 
\bar{P}_{t} 
+ \frac{\alpha'}{2} \bigl(\frac{\alpha' u}{4}\bigr) [P_{t}(\bar{C}_{u}-\bar{C}_{s}) + \bar{P}_{t}(C_{u}-C_{s})]\nonumber\\
&& + \bigl(\frac{\alpha'}{2}\bigr)^{2}(C_{u}-C_{s})(\bar{C}_{u}-\bar{C}_{s})\Bigr]= \tilde{G}_{40t} \equiv \frac{B \tilde{H}_{t}}{\frac{\alpha' t}{4}}.\label{A5}
\end{eqnarray} 
\par Similarly, using the property $(C_{u}-C_{t})(1 \rightarrow 4)=C_{s}-C_{t}$, one learns that
\begin{eqnarray}
&& \tilde{G}_{40s}(1\rightarrow 4)= \frac{B}{\frac{\alpha' u}{4}} \Bigl[\bigl(\frac{\alpha' s}{4}\bigr)^{2} P_{u} 
\bar{P}_{u} 
+ \frac{\alpha'}{2} \bigl(\frac{\alpha' s}{4}\bigr) [P_{u}(\bar{C}_{s}-\bar{C}_{t}) + \bar{P}_{u}(C_{s}-C_{t})]\nonumber\\
&& + \bigl(\frac{\alpha'}{2}\bigr)^{2}(C_{s}-C_{t})(\bar{C}_{s}-\bar{C}_{t})\Bigr]= \tilde{G}_{40u} \equiv \frac{B \tilde{H}_{u}}{\frac{\alpha' u}{4}}.\label{A6}
\end{eqnarray}  
\par Furthermore, using the property $(C_{u}-C_{s})(3 \rightarrow 4)=C_{t}-C_{s}$ it is easy to show that 
$\tilde{G}_{40t}(3 \rightarrow 4)=\tilde{G}_{40u}$ at the residue of the pole $u=0$. This completes the proof that at the residue 
of the massless pole in the appropriate channel, the amplitudes in the massless gravitational sector are $stu$ symmetric.  
\par We investigate now the $stu$ symmetry of the scattering amplitudes in the tachyon sector. Again, the 
terms with potential violation of the $stu$ symmetry come from the $\bar{C}_{s}, \bar{C}_{t}$ and $\bar{C}_{u}$ 
contributions. In the tardyon sector a similar reasoning is valid but with their left-mover sector counterparts. In the $s$-channel the term 
to be analyzed is just 
\begin{eqnarray}
&& \tilde{g}_{4s-} = \frac{\alpha'}{2} P_{s} \Bigl[\frac{\frac{\alpha' t}{4} \bar{C}_{t} + \frac{\alpha' u}{4} \bar{C}_{u} - \bar{C}_{s}}{1 + \frac{\alpha' s}{4}}\Bigr], \label{A7}
\end{eqnarray}
since all other terms in the $s$-channel of the tachyon sector contained in the scattering amplitude are explicitly $stu$ symmetric. In order 
to develop this expression, we use the kinematic constraint $s+t+u=0$ and write $u=-\frac{s}{2}+\frac{(u-t)}{2}$ and 
$t=-\frac{s}{2}-\frac{(u-t)}{2}$. Thus, the bracket can be rewritten as
\begin{eqnarray}
&& \frac{\alpha' t}{4} \bar{C}_{t} + \frac{\alpha' u}{4} \bar{C}_{u} - \bar{C}_{s} = - \frac{\alpha' s}{8}(\bar{C}_{t} + \bar{C}_{u}) 
- \bar{C}_{s} + \frac{\alpha'}{8}(u-t)(\bar{C}_{u} - \bar{C}_{t}).\label{A8}
\end{eqnarray} 
When this is computed at the tachyon pole value, we find
\begin{eqnarray}
&& \Bigl[\frac{\alpha' t}{4} \bar{C}_{t} + \frac{\alpha' u}{4} \bar{C}_{u} - \bar{C}_{s}\Bigr]_{\frac{\alpha' s}{4}=-1} = \frac{1}{2}(\bar{C}_{t} + \bar{C}_{u} - 2\bar{C}_{s}) + \frac{\alpha'}{8}(u-t)(\bar{C}_{u} - \bar{C}_{t}).\label{A9}
\end{eqnarray}
\par Care must be exercised when performing the transposition here, since it takes the amplitude into one channel and leads to 
its value in another channel. In a particular channel, the Mandelstam variable corresponding to that channel is fixed at the value of the pole 
of the amplitude when we compute its residue. For example, in the $s$-channel that amplitude is computed at the tachyon pole value 
$\frac{\alpha' s}{4}=-1$. The possible transpositions starting from the $s$-channel take the residue of the amplitude to that in the 
$t$-channel (where now $t$ is fixed at $\frac{\alpha' t}{4}=-1$) or to the one in the $u$-channel ($u$ is fixed at $\frac{\alpha' u}{4}=-1$). In 
other words, the difference in the Mandelstam variables coming from the second term in last equation also gets transformed.
\par Taking the transposition $1 \rightarrow 3$ in last equation we find
\begin{eqnarray}
&& \Bigl[\frac{\alpha' t}{4} \bar{C}_{t} + \frac{\alpha' u}{4} \bar{C}_{u} - \bar{C}_{s}\Bigr]_{\frac{\alpha' s}{4}=-1} (1 \rightarrow 3) = 
\frac{1}{2}(\bar{C}_{t} + \bar{C}_{u} - 2\bar{C}_{s})(1 \rightarrow 3) + \frac{\alpha'}{8}(u-s)\;\; \times\nonumber\\
&& (\bar{C}_{u} - \bar{C}_{t})(1 \rightarrow 3).\label{A10}
\end{eqnarray}
\par Note that $(\bar{C}_{0t} + \bar{C}_{0u} - 2\bar{C}_{0s})(1 \rightarrow 3)=\bar{C}_{0t} + \bar{C}_{0s} - 2\bar{C}_{0t}$. Using the resources 
presented so far its is easy to see that $(\bar{C}_{1u} + \bar{C}_{1t})(1 \rightarrow 3)= \bar{C}_{1u} - 2\bar{C}_{1t}$. This implies that
$(\bar{C}_{t} + \bar{C}_{u} - 2\bar{C}_{s})(1 \rightarrow 3)= (\bar{C}_{s} + \bar{C}_{u} - 2\bar{C}_{t})$. The arguments just presented lead us 
to conclude that
\begin{eqnarray}
&& \Bigl[\frac{\alpha' t}{4} \bar{C}_{t} + \frac{\alpha' u}{4} \bar{C}_{u} - \bar{C}_{s}\Bigr]_{\frac{\alpha' s}{4}=-1} (1 \rightarrow 3) = 
\frac{1}{2}(\bar{C}_{s} + \bar{C}_{u} - 2\bar{C}_{t}) + \frac{\alpha'}{8}(u-s)(\bar{C}_{u} - \bar{C}_{s}).\label{A11}
\end{eqnarray} 
\par In the $t$-channel, employ variables $s=-\frac{t}{2}-\frac{(u-s)}{2}$ and $u=-\frac{t}{2}+\frac{(u-s)}{2}$. By the same reasoning, the 
residue of the quantity 
\begin{eqnarray}
&& \frac{\alpha' s}{4} \bar{C}_{s} + \frac{\alpha' u}{4} \bar{C}_{u} - \bar{C}_{t} = - \frac{\alpha' t}{8}(\bar{C}_{s} + \bar{C}_{u}) 
- \bar{C}_{t} + \frac{\alpha'}{8}(u-s)(\bar{C}_{u} - \bar{C}_{s}),\label{A12}
\end{eqnarray}
appearing in the expression of the amplitude in $t$-channel, at the tachyon pole value $\frac{\alpha' t}{4}=-1$ is 
$\Bigl(\frac{\alpha' s}{4} \bar{C}_{s} + \frac{\alpha' u}{4} \bar{C}_{u} - \bar{C}_{t}\Bigl)_{\frac{\alpha' t}{4}=-1}=\Bigl[\frac{\alpha' t}{4} \bar{C}_{t} + \frac{\alpha' u}{4} \bar{C}_{u} - \bar{C}_{s}\Bigr]_{\frac{\alpha' s}{4}=-1} (1 \rightarrow 3)$. Considering $\tilde{g}_{4s-}$ and making the transposition $1 \rightarrow 3$, we find:
\begin{eqnarray}
&& \tilde{g}_{4s-} (1 \rightarrow 3) = \frac{\alpha'}{2} P_{t} \frac{\Bigl[\frac{\alpha' s}{4} \bar{C}_{s} + \frac{\alpha' u}{4} \bar{C}_{u} - \bar{C}_{t}\Bigr]_{\frac{\alpha' t}{4}=-1}}{1 + \frac{\alpha' t}{4}} \equiv \tilde{g}_{4t-}, \label{A13}
\end{eqnarray}
which constitutes an explicit proof of the $s \rightarrow t$ symmetry (and vice-versa) of the scattering amplitude under the permutation 
$1 \rightarrow 3$.
\par Continue in the $s$-channel and now take the transposition $1 \rightarrow 4$ which changes $s \rightarrow u$ or in other words, takes 
amplitudes in the $s$-channel and transforms them into $u$-channel ones. Perform this transposition in Eq. (\ref{A9}) in order to obtain the 
preliminary result 
\begin{eqnarray}
&& \Bigl[\frac{\alpha' t}{4} \bar{C}_{t} + \frac{\alpha' u}{4} \bar{C}_{u} - \bar{C}_{s}\Bigr]_{\frac{\alpha' s}{4}=-1} (1 \rightarrow 4) = 
\frac{1}{2}(\bar{C}_{t} + \bar{C}_{u} - 2\bar{C}_{s})(1 \rightarrow 4) + \frac{\alpha'}{8}(s-t)\;\;\;\times\nonumber\\
&& (\bar{C}_{u} - \bar{C}_{t})(1 \rightarrow 4).\label{A14}
\end{eqnarray}
\par By noting that $\bar{C}_{1t}+\bar{C}_{1u}(1 \rightarrow 4) = \bar{C}_{1t}-2\bar{C}_{1u}$ and using previous results already given for the 
transposition, one can show that
\begin{eqnarray}
&& \Bigl[\frac{\alpha' t}{4} \bar{C}_{t} + \frac{\alpha' u}{4} \bar{C}_{u} - \bar{C}_{s}\Bigr]_{\frac{\alpha' s}{4}=-1}(1 \rightarrow 4) = 
\frac{1}{2}(\bar{C}_{t} + \bar{C}_{s} - 2\bar{C}_{u}) + \frac{\alpha'}{8}(s-t) \;\;\;\times\nonumber\\
&& (\bar{C}_{s} - \bar{C}_{t}).\label{A15}
\end{eqnarray}
\par In the $u$-channel, use variables $s=-\frac{u}{2}+ \frac{(s-t)}{2}$ and $t=-\frac{u}{2} - \frac{(s-t)}{2}$ in order to write the 
combination below at the tachyon pole as
\begin{eqnarray}
&& \Bigl[\frac{\alpha' s}{4} \bar{C}_{s} + \frac{\alpha' t}{4} \bar{C}_{t} - \bar{C}_{u}\Bigr]_{\frac{\alpha' u}{4}=-1} = 
\frac{1}{2}(\bar{C}_{t} + \bar{C}_{s} - 2\bar{C}_{u}) + \frac{\alpha'}{8}(s-t) (\bar{C}_{s} - \bar{C}_{t}).\label{A16}
\end{eqnarray}
\par It is now obvious that
\begin{eqnarray}
&& \tilde{g}_{4s-} (1 \rightarrow 4) = \frac{\alpha'}{2} P_{u} \frac{\Bigl[\frac{\alpha' s}{4} \bar{C}_{s} + \frac{\alpha' t}{4} \bar{C}_{t} - \bar{C}_{u}\Bigr]_{\frac{\alpha' u}{4}=-1}}{1 + \frac{\alpha' u}{4}} \equiv \tilde{g}_{4u-}, \label{A17}
\end{eqnarray}
and the $s \rightarrow u$ symmetry under the exchange $1 \rightarrow 4$ is proved. 
\par The $t \rightarrow u$ symmetry under the permutation $3 \rightarrow 4$ is simpler to prove. The transposition in the terms which are 
$O(\alpha^{' 0})$ is trivial as before. The residue of Eq. (\ref{A12}) in the $t$-channel taken along with the transposition produces the preliminary result
\begin{eqnarray}
&& \Bigl[\frac{\alpha' s}{4} \bar{C}_{s} + \frac{\alpha' u}{4} \bar{C}_{u} - \bar{C}_{t}\Bigr]_{\frac{\alpha' t}{4}=-1} (3 \rightarrow 4) = \frac{1}{2}(\bar{C}_{s} + \bar{C}_{u} -2 \bar{C}_{t}) (3 \rightarrow 4)\nonumber\\
&&  + \frac{\alpha'}{8}(t-s)(\bar{C}_{u} - \bar{C}_{s})(3 \rightarrow 4),\label{A18}
\end{eqnarray}
\par It is a simple task to prove that $(\bar{C}_{s} + \bar{C}_{u} -2 \bar{C}_{t}) (3 \rightarrow 4)= (\bar{C}_{s} + \bar{C}_{t} -2 \bar{C}_{u})$. 
Since $\bar{C}_{1u}(3 \rightarrow 4)=\bar{C}_{1t}$, we get to
\begin{eqnarray}
&& \Bigl[\frac{\alpha' s}{4} \bar{C}_{s} + \frac{\alpha' u}{4} \bar{C}_{u} - \bar{C}_{t}\Bigr]_{\frac{\alpha' t}{4}=-1} (3 \rightarrow 4) = 
\frac{1}{2}(\bar{C}_{s} + \bar{C}_{t} - 2 \bar{C}_{u}) + \frac{\alpha'}{8}(t-s) \;\;\;\times\nonumber\\
&& (\bar{C}_{t} - \bar{C}_{s}).\label{A19}
\end{eqnarray}
\par The piece of the scattering amplitude when the tachyon is the intermediate state in the $t$-channel is given by Eq. (\ref{A13}). Performing the 
transposition and using the above result in conjumination with Eqs. (\ref{A16}) and (\ref{A17}) produces the desired $t \rightarrow u$ symmetry 
$\tilde{g}_{4t-}(3 \rightarrow 4)= \tilde{g}_{4u-}$. This concludes the explicit proof of the $stu$ symmetry for the residue of the 
scattering amplitude in tachyon sector.
\par Finally, let us consider the tardyon sector in the $s$-channel. Again, here we concentrate only in the piece of the amplitude which depends 
only on $C_{s}, C_{t}$ and $C_{u}$. Bearing in mind that these quantities inherit the same properties of their right-mover side counterparts, 
and employing the variables $u=-\frac{s}{2}+\frac{(u-t)}{2}$ and $t=-\frac{s}{2}-\frac{(u-t)}{2}$ the appropriate combination of these 
objects in the tardyon amplitude reads
\begin{eqnarray}
&& \frac{\alpha' t}{4} C_{t} + \frac{\alpha' u}{4} C_{u} +  C_{s} = -\frac{\alpha' s}{8}(C_{t}+C_{u}) + C_{s} + \frac{\alpha'}{8}(u-t)(C_{u}-C_{t}).\label{A20}
\end{eqnarray}
\par The residue of this expression at the tardyon pole $\frac{\alpha' s}{4}=1$ is
\begin{eqnarray}
&& \Bigl[\frac{\alpha' t}{4}C_{t} + \frac{\alpha' u}{4}C_{u} +  C_{s}\Bigr]_{\frac{\alpha' s}{4}=1}= -\frac{1}{2}(C_{t}+C_{u} -2 C_{s}) 
+ \frac{\alpha'}{8}(u-t)(C_{u}-C_{t}). \label{A21} 
\end{eqnarray}
\par From last equation, one can see that the combination of these coefficients at the 
tardyon pole is identical to that appearing in the tachyon pole in terms of the coordinate $u-t$. Define the residues of the part of the 
amplitude involving only these terms (with a tardyon as an intermediate state in the $s,t$ and $u$ 
channels) as   
\begin{subequations}
\begin{eqnarray}
&& \tilde{g}_{4s+} = -\frac{\alpha'}{2} \frac{\bar{P}_{s}}{1+\frac{\alpha' s}{4}}\Bigl[\frac{\alpha' t}{4}C_{t} + \frac{\alpha' u}{4}C_{u} +  C_{s}\Bigr]_{\frac{\alpha' s}{4}=1};\label{A22a}\\
&& \tilde{g}_{4t+} = -\frac{\alpha'}{2} \frac{\bar{P}_{t}}{1+\frac{\alpha' t}{4}}\Bigl[\frac{\alpha' s}{4}C_{s} + \frac{\alpha' u}{4}C_{u} +  C_{t}\Bigr]_{\frac{\alpha' t}{4}=1};\label{A22b}\\
&& \tilde{g}_{4u+} = -\frac{\alpha'}{2} \frac{\bar{P}_{u}}{1+\frac{\alpha' u}{4}}\Bigl[\frac{\alpha' s}{4}C_{s} + \frac{\alpha' t}{4}C_{t} +  C_{u}\Bigr]_{\frac{\alpha' u}{4}=1}.\label{A22c}
\end{eqnarray}
\end{subequations}
\par From the form of the residues of the amplitudes in different channels as written in terms of variables $(u-t)$ ($s$-channel), $(u-s)$ 
($t$-channel) and $(s-t)$ ($u$-channel) and our analysis of the tachyon sector, it follows trivially that $\tilde{g}_{4s+}(1 \rightarrow 3)=\tilde{g}_{4t+}$, $\tilde{g}_{4s+}(1 \rightarrow 4)=\tilde{g}_{4u+}$ and $\tilde{g}_{4t+}(3 \rightarrow 4)=\tilde{g}_{4u+}$, concluding the explicit 
proof of the $stu$ symmetry in all sectors.

\end{document}